\documentclass[aps,prx,twocolumn,superscriptaddress,nofootinbib,floatfix]{revtex4-1}
 \usepackage[utf8]{inputenc}
\usepackage{color}
\usepackage{tikz}
\usepackage{float}

\usepackage{lipsum}

\usepackage{upgreek}
\usepackage[normalem]{ulem}
\usepackage{mathtools}
\usepackage{datetime}

\usepackage{amsmath}
\usepackage{amssymb}

\usepackage{tikz}
\usepackage{amsthm,euscript}

\usepackage{ragged2e} 
\newif\ifmarkreview
\markreviewfalse   

\ifmarkreview
  \newcommand{\rev}[1]{{\color{red}#1}}
\else
  \newcommand{\rev}[1]{{\color{black}#1}}
\fi

\usepackage{mathptmx}

\usepackage[bbgreekl]{mathbbol} 
\DeclareMathAlphabet{\mathcal}{OMS}{cmsy}{m}{n}
\renewcommand{\mathcal}[1]{\text{\usefont{OMS}{cmsy}{m}{n}#1}}

\usepackage[all,cmtip,knot]{xy}
\usepackage{xypic}
\usepackage{braids}

\usepackage[bbgreekl]{mathbbol} 

\usepackage[all,cmtip,knot]{xy} 
\usepackage{mathrsfs}
\usepackage[unicode=true,bookmarks=true,bookmarksnumbered=false,bookmarksopen=false,breaklinks=false,pdfborder={0 0 1},backref=false,colorlinks=true]{hyperref}
\usepackage{algorithm}
\usepackage{algorithmic}

\definecolor{quantumblue}{HTML}{002366} 
\definecolor{quantumgreen}{HTML}{007474} 
\definecolor{quantumgray}{HTML}{555555} 

\hypersetup{
    colorlinks,
    linkcolor={gray!80!black},
    citecolor={blue!90!black},
    urlcolor={gray!60!black}
}

\makeatletter

\usepackage{graphicx}
\usepackage{epstopdf}

\renewcommand{\vec}[1]{\boldsymbol{#1}}

\newcommand{\orcid}[1]{\href{https://orcid.org/#1}{\includegraphics[width=8pt]{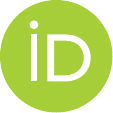}}}

\makeatletter
\newsavebox{\@brx}
\newcommand{\llangle}[1][]{\savebox{\@brx}{\(\m@th{#1\langle}\)}%
  \mathopen{\copy\@brx\kern-0.5\wd\@brx\usebox{\@brx}}}
\newcommand{\rrangle}[1][]{\savebox{\@brx}{\(\m@th{#1\rangle}\)}%
  \mathclose{\copy\@brx\kern-0.5\wd\@brx\usebox{\@brx}}}
\makeatother

\theoremstyle{definition}

\theoremstyle{remark}

\newcommand\Tr   {\operatorname{Tr}}

\newcommand{\identity}{\mathds{I}}

\usepackage{bm,latexsym,galois,euscript,dsfont}

\newcommand{\valpha}{\vec{\alpha}}
\newcommand{\vbeta}{\vec{\beta}}

\usepackage{tikz}
\usetikzlibrary{positioning,intersections}
\usetikzlibrary{calc}
\usetikzlibrary{shapes.geometric}
\usepackage{braids}
\usepackage{tikz-cd}

\definecolor{nblue}{rgb}{0.2,0.2,0.7}
\definecolor{ngreen}{rgb}{0.2,0.6,0.2}
\definecolor{nred}{rgb}{0.7,0.2,0.2}
\definecolor{nblack}{rgb}{0,0,0}

\makeatother

\begin{document}

\title{Variational Transformer Ansatz for the Density Operator of Steady States in Dissipative Quantum Many-Body Systems}

\author{Lu Wei}

\affiliation{Department of Applied Mathematics \& Statistics, Stony Brook University, Stony Brook, NY 11794, USA}
\affiliation{Program of Data Science, Stony Brook University, Stony Brook, NY 11794, USA}

\author{Zhian Jia\orcid{0000-0001-8588-173X}}
\email{giannjia@foxmail.com}
\affiliation{Centre for Quantum Technologies, National University of Singapore, SG 117543, Singapore}
\affiliation{Department of Physics, National University of Singapore, SG 117543, Singapore}

\author{Yufeng Wang\orcid{0000-0003-0640-8536}}
\affiliation{Department of Computer Science, Stony Brook University, Stony Brook, NY 11794, USA}

\author{Dagomir Kaszlikowski}
\affiliation{Centre for Quantum Technologies, National University of Singapore, SG 117543, Singapore}
\affiliation{Department of Physics, National University of Singapore, SG 117543, Singapore}

\author{Haibin Ling}
\email{linghaibin@westlake.edu.cn}
\affiliation{Department of Artificial Intelligence, Westlake University, 
        Hangzhou, Zhejiang 310030, China}

\begin{abstract}

The transformer architecture, known for capturing long-range dependencies and intricate patterns, has extended beyond natural language processing. Recently, it has attracted significant attention in quantum information and condensed matter physics. In this work, we propose the \textit{transformer density operator ansatz} for determining the steady states of dissipative quantum many-body systems. By vectorizing the density operator as a many-body state in a doubled Hilbert space, the transformer encodes the amplitude and phase of the state's coefficients, with its parameters serving as variational variables. Our design preserves translation invariance while leveraging attention mechanisms to capture diverse long-range correlations. We demonstrate the effectiveness of our approach by numerically calculating the steady states of dissipative Ising and Heisenberg spin chain models, showing that our method achieves excellent accuracy in predicting mixed steady states.

\end{abstract}
\maketitle

\emph{Introduction.} ---
\noindent The investigation of open quantum systems has experienced a surge in interest in recent years. 
From a fundamental perspective, despite significant experimental strides in isolating quantum systems, a finite coupling to the environment is unavoidable, imparting dynamic characteristics that encompass a diverse range of features not observed in equilibrium systems \cite{breuer2002theory,Weimer2021simulation}. In practical terms, these systems offer a platform for employing controlled dissipation channels to engineer captivating quantum states as the stationary outcome of their dynamics, thus holding potential applications in quantum information tasks \cite{gardiner2004quantum,Diehl2008,Verstraete2009}.
Diverging from closed quantum systems, where a wave function is commonly used to represent the quantum state, the focus of study in open quantum systems shifts to the density operator $\rho$. Effectively describing interacting open quantum many-body systems presents a significant challenge for both theoretical and numerical approaches~\cite{Weimer2021simulation}.

The evolution of an open quantum system is governed by the master equation, and several methods have been developed to solve it in recent years. These include analytic approaches based on the Keldysh formalism~\cite{sieberer2016keldysh}, tensor network techniques such as the density matrix renormalization group and matrix product operator methods~\cite{Schollwock2005DMRG, orus2018tensor, Cirac2021MPSreivew, Cui2015MPO, kshetrimayum2017simple, Werner2016positive}, the quantum-trajectory method~\cite{molmer93, vicentini2019stochastic}, the cluster mean-field approach~\cite{Jin2016cluster}, phase space methods~\cite{luo2022autoregressive}, and corner-space renormalization~\cite{Finazzi2015corner}, among others.

\begin{figure}[b]
\centering
\includegraphics[width=0.7\linewidth]{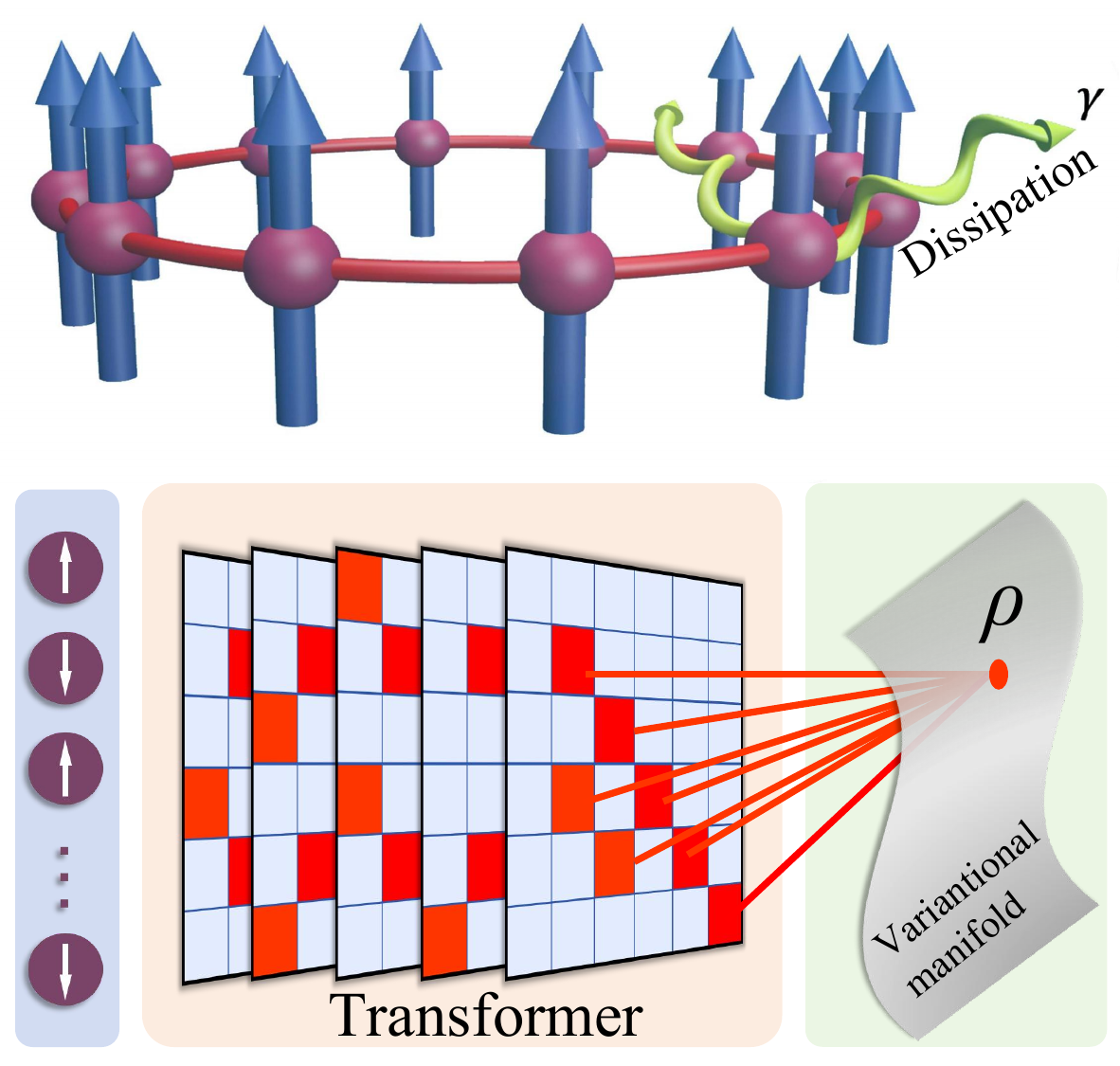}
\caption{\label{fig:ising} Illustration of a dissipative spin chain with periodic boundary condition (top) and its transformer representation of the density operator (bottom). The dissipative rate $\gamma$ describes the strength of the coupling to the environment, which leads to decoherence and information loss in the system.}
\end{figure}

Variational methods are fundamental in the study of quantum many-body systems, offering deep insights into the properties of highly complex physical systems. Neural networks have the capacity to efficiently extract hidden patterns from large datasets~\cite{bishop1995neural, fausett1994fundamentals}. In recent years, neural network-based variational ansatz states have garnered significant attention for solving quantum problems, see \textit{e.g.}~\cite{jia2019quantum, Carleo2019machine,jia2018efficient,jia2020entanglement,gao2017efficient,Deng2017,zhang2022efficientRBM}. 
The most well-studied examples are Restricted Boltzmann Machine (RBM) states~\cite{Carleo602}. Beyond RBMs, other architectures, such as deep Boltzmann machines, convolutional neural networks (CNN), and feedforward neural networks have also been employed to construct neural network ansatz states.
Many of these neural network ansatz methods have been extended to open quantum systems~\cite{Torlai2018latent, Yoshioka2019neural, Hartmann2019neural, Nagy2019variational, vicentini2019variational, Nigro2021nns,mellak2024deep,Kothe2024NNDO}, where density operators are encoded into neural networks. \rev{However, most existing neural quantum state approaches for open systems rely on architectures with inherent locality constraints, limiting their effectiveness in capturing the complex, long-range correlated mixed states that commonly arise as steady states of dissipative systems}.

\begin{figure*}[ht]
\centering
        \includegraphics[width=0.9\linewidth]{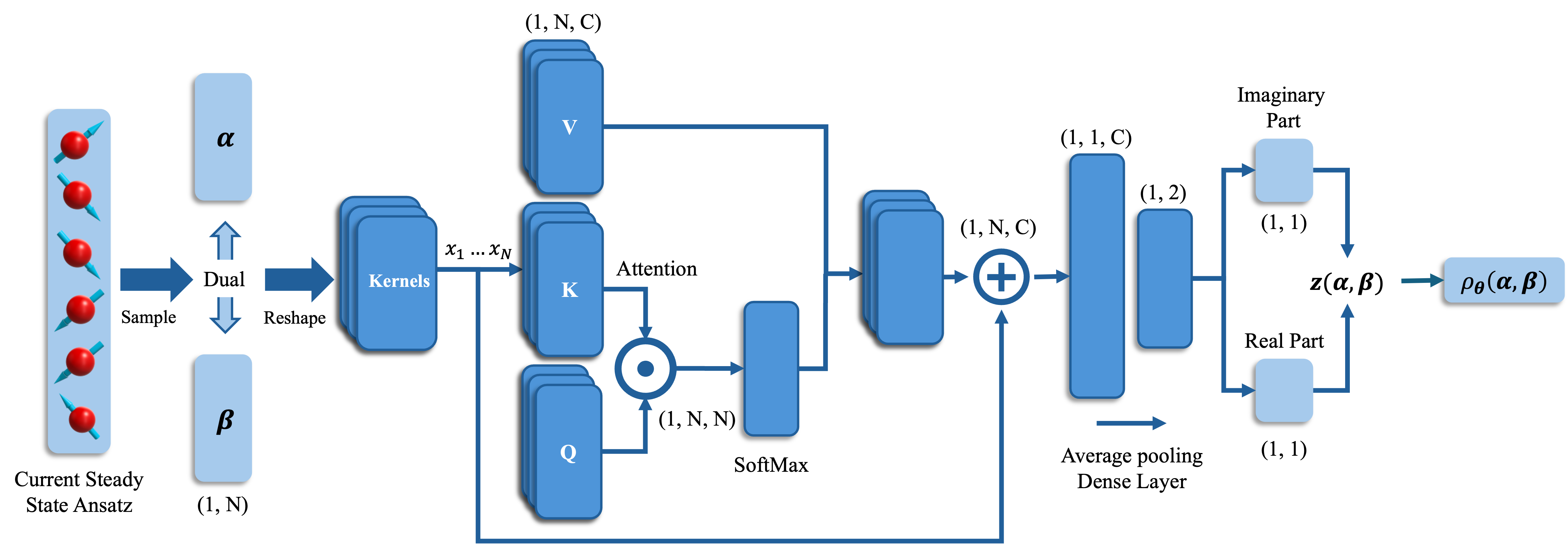}
    \caption{Schematic representation of the transformer density operator ansatz for the steady-state density operator of an open quantum spin chain with periodic boundary conditions. The input spin configurations \((\vec{\alpha}, \vec{\beta})\) are split into left and right components and then stacked and reshaped to form the input to two convolutional layers with circular padding, which embed the local features. A self-attention block then captures long-range dependencies by allowing each spin site to attend to all others through learned attention weights. Global average pooling ensures translation invariance, followed by a fully connected layer that maps the spatially averaged vectors into the real and imaginary parts of a complex output. The last step symmetrizes this output to enforce the Hermiticity of the density operator. For the experiments on both Ising and Heisenberg chains presented in this paper, we employ the exact same architecture.
    }
\label{fig:Transformer}
\end{figure*}

The transformer architecture has recently gained significant attention due to its success in natural language processing tasks~\cite{vaswani2023attentionneed}. It has also been successfully applied to many-body problems in closed quantum systems~\cite{Zhang2023Transformer, viteritti2023transformer,rende2025networkquantumstates}. However, its application as a steady state ansatz for solving open quantum systems remains relatively underexplored. \rev{The key advantage of transformers lies in their self-attention mechanism, which can efficiently capture arbitrary long-range correlations without the locality constraints inherent in tensor network methods. This makes them particularly well-suited for open quantum systems, where dissipation can induce complex, non-local correlation patterns in the steady state}.

In Ref.~\cite{luo2022autoregressive}, the quantum state is mapped to a probability distribution in phase space, and its evolution is reformulated as a probabilistic equation, enabling the transformer to simulate open quantum system dynamics. In this work, we introduce the \textit{transformer density operator ansatz}, based on the vectorization of the density operator—an approach that has recently gained attention in studies of open-system quantum phases, weak and strong symmetries, tenfold classification, and related topics (see, e.g., \cite{Kawabata2023symmetry,bao2023mixedstate,Sohal2025noisy,Yoshioka2019neural,ma2024SPTmixed,Kothe2024NNDO}). We employ this ansatz variationally to solve for the steady state of dissipative quantum systems.
As we will demonstrate using the dissipative spin chain model, this approach can efficiently capture the steady state with high precision.

\vspace{1em}
\emph{Transformer density operator ansatz.} ---
Consider an $N$-particle system with the Hilbert space $\mathcal{H}$ spanned by the basis states $|\vec{\alpha}\rangle$, where $\vec{\alpha} = (\alpha_1, \dots, \alpha_N)$ labels the states of the $N$ degrees of freedom composing the system. For example, in a qubit system, $\alpha_i$ take values in \(\{0, 1\}\). For a density operator $\rho \in B(\mathcal{H})$ (where $B(\mathcal{H})$ denotes the space of all linear operators), which is a positive semidefinite, trace-one and Hermitian operator, we can express its matrix elements as $\rho(\vec{\alpha}, \vec{\beta}) = \langle \vec{\alpha} | \rho | \vec{\beta} \rangle$ in the basis of $|\vec{\alpha}\rangle$ and $|\vec{\beta}\rangle$. By vectorization, $\rho$ can be transformed into a vector $|\rho \rrangle = \sum_{\vec{\alpha}, \vec{\beta}} \rho(\vec{\alpha}, \vec{\beta}) |\vec{\alpha}\rangle |\vec{\beta}\rangle$ in the doubled Hilbert space $\mathcal{H} \otimes \mathcal{H}$ (see supplementary material for further details). 
In order to construct a variational transformer representation of the density operator, we express the vectorized density operator as
\begin{align}
|\rho_{\bm{\theta}}(\mathbf{J})\rrangle
&= \sum_{\vec{\alpha},\vec{\beta}} \rho_{\bm{\theta}}(\vec{\alpha},\vec{\beta},\mathbf{J}) |\vec{\alpha}\rangle |\vec{\beta}\rangle,
\label{eq:transformer_cnn_rho}
\end{align}
where \(\rho_{\bm{\theta}} (\vec{\alpha},\vec{\beta},\mathbf{J})\) is the density operator's complex amplitude corresponding to the configuration of $(\vec{\alpha}, \vec{\beta})$. The variational parameters ${\bm{\theta}}$ define the model, and $\mathbf{J}$ denotes the physical parameters of the open quantum system, which will be ignored for simplicity in the following description. 

Our transformer density operator ansatz is mainly parameterized by incorporating convolutional layers for local feature extraction and the transformer block with a self-attention mechanism to capture long-range correlations within the density matrix structure. Specifically, our ansatz is entirely parametrized by real-valued parameters, and the final complex output is obtained by combining two real-valued outputs that represent its real and imaginary components. For illustration in Fig.~\ref{fig:Transformer}, we set the batch number to 1 in the schematic, with additional details provided in Sec.~\ref{sec:TansformerDOA} of the supplementary material.

Our goal is to compute $\rho_{\vec{\theta}} (\vec{\alpha},\vec{\beta})$ for each configuration $(\vec{\alpha}, \vec{\beta})$. This requires sampling from the configuration space, as detailed in Sec.~\ref{SM:sample} of the supplementary material. Below, we provide a step-by-step discussion on obtaining the steady state.

The sampled input \((\vec{\alpha}, \vec{\beta})\) is first reshaped and passed through two convolutional layers that serve as a feature encoding stage. In this stage, the input configuration is convolved with a bank of learnable convolutional filters, each with a kernel size of two‐by‐one, to encode local feature embeddings. To preserve the periodic boundary conditions of the system, circular padding is applied in the convolutional layers, ensuring that the first and last sites are treated equivalently. Each convolutional operation is followed by a nonlinear activation function. This process gives us a set of feature vectors \(\{\vec{x}_1, \dots, \vec{x}_i, \dots, \vec{x}_N\}\), where \(\vec{x}_i\) represents an embedded feature for the \(i\)-th spin.

Subsequently, these local feature embeddings are processed by a transformer encoder block~\cite{vaswani2023attentionneed} with the self-attention module to capture long-range correlations that can emerge globally in a strongly interacting Ising chain. We introduce three learnable matrices \(Q\), \(K\), and \(V\), each of which transforms an input feature vector into a corresponding query, key, and value representation. Concretely, for any feature vector \(\vec{x}_i\), we define
\begin{equation}
    \vec{q}_i = Q\,\vec{x}_i,\quad
    \vec{k}_i = K\,\vec{x}_i,\quad
    \vec{v}_i = V\,\vec{x}_i,
\end{equation}
where \(Q\), \(K\), and \(V\) share the same shape but are learned independently to capture different aspects of the input features. The attention mechanism $\omega(\cdot,\cdot)$ allows each site to attend to all other sites by computing the learned attention weights using a scaled dot product, followed by a softmax operation
\begin{equation}  
\omega(\vec{q}_i, \vec{k}_j) = \frac{\exp\left( \frac{\langle \vec{q}_i, \vec{k}_j \rangle}{\sqrt{d}} \right)}{\sum_{j=1}^N \exp\left( \frac{\langle \vec{q}_i, \vec{k}_j \rangle}{\sqrt{d}} \right)},
\end{equation}
where \(d\) is the dimension of the query, key, and value vectors, $\vec{q}_i, \vec{k}_j, \vec{v}_j \in \mathbb{R}^d$. Dividing by \(\sqrt{d}\) keeps the dot-product within a more stable numeric range, preventing large vector sizes from causing the exponential function to overflow. The attention weights \(\omega\) measure how much the $j$-th input should contribute to the $i$-th context vector. Using these attention weights, the context vector for each site is constructed as
\begin{equation}
    \vec{a}_i = \sum_{j=1}^N \omega(\vec{q}_i, \vec{k}_j)\,\vec{v}_j.
\end{equation}
The context vectors \(\{\vec{a}_1, \dots, \vec{a}_N\}\) encode global correlations across the entire system. The output context vectors ${\vec{a}_1, \dots, \vec{a}_N}$ are computed in parallel and added with feature vectors \(\{\vec{x}_1, \dots, \vec{x}_N\}\). To further enhance the model’s capacity to capture diverse interactions, the self-attention mechanism is extended to multi-head attention. In this setting, the feature channels are split into \(m\) heads, with independent sets of query, key, and value matrices \(Q^\mu\), \(K^\mu\), and \(V^\mu\) (for \(\mu = 1, \dots, m\)) applied to each head; the outputs from all heads are then concatenated to form the final representation.

The attention mechanism enables the network to capture arbitrary pairwise relationships, which is particularly beneficial in open quantum systems where dissipation and quantum coherence can induce correlations of long-range neighbors. Furthermore, with the multi-head attention, the ansatz may exhibit multiple distinct correlation stereotypes since different heads can specialize in capturing different scales of correlation.

To ensure translation invariance in the final output, which is crucial for homogeneous spin systems under periodic boundary condition, we average over the positions in the chain $\vec{h} = \frac{1}{N} \sum_{i=1}^N \left( \vec{a}_i + \vec{x}_i \right)$. 
This eliminates explicit dependence on site indices and dramatically reduces the number of free parameters in the subsequent layer.

The mean-pooled vector \(\vec{h}\) is then fed into a fully-connected layer that produces two real-valued outputs, which correspond to the real part and the imaginary part of a complex number $z(\vec{\alpha},\vec{\beta})$. To ensure Hermiticity, the final representation of the steady-state density matrix element \(\rho_{\vec{\theta}} (\vec{\alpha},\vec{\beta})\) is obtained by symmetrizing the previous complex output:
\begin{equation}
\label{eq:log_sum_exp_conjugate}
\rho_{\bm \theta}(\vec{\alpha},\vec{\beta})=\log \Bigl(\exp \bigl[z(\vec{\alpha},\vec{\beta})\bigr] 
         + \exp \bigl[z(\vec{\beta},\vec{\alpha})\bigr]^{*}\Bigr).
\end{equation}
This transformation ensures that the resulting quantity satisfies $\rho_{\bm \theta}(\vec{\alpha},\vec{\beta}) = \rho_{\bm \theta}(\vec{\beta},\vec{\alpha})^{*}$ and also maintains numerical stability. However, positive semidefiniteness of the density is not explicitly guaranteed and is instead learned through optimization as described in~\cite{vicentini2022netket}. Now, for a configuration pair $(\valpha,\vbeta)$, we are able to give the corresponding complex amplitude $\rho_{\bm \theta}(\vec{\alpha},\vec{\beta})$ of its steady state in Eq.~\eqref{eq:transformer_cnn_rho} through the transformer density operator ansatz. 

To summarize, our transformer density operator ansatz uses convolutional filters with circular padding for local encoding and multi-head self-attention to capture long-range correlations. Global pooling enforces translation invariance, and a subsequent symmetrization ensures Hermiticity.
By parameterizing the complex amplitude \(\rho_{\theta}(\vec{\alpha},\vec{\beta})\) in Eq.~\eqref{eq:transformer_cnn_rho} with transformer density operator ansatz, we are able to maintain the essential properties of Hermiticity and approximate positivity. 
The self-attention mechanism enables the ansatz to effectively capture intricate correlation patterns inherent in open quantum systems and scale efficiently to larger spin chains.

\begin{figure*}[th]
\centering
\includegraphics[width=1\textwidth]{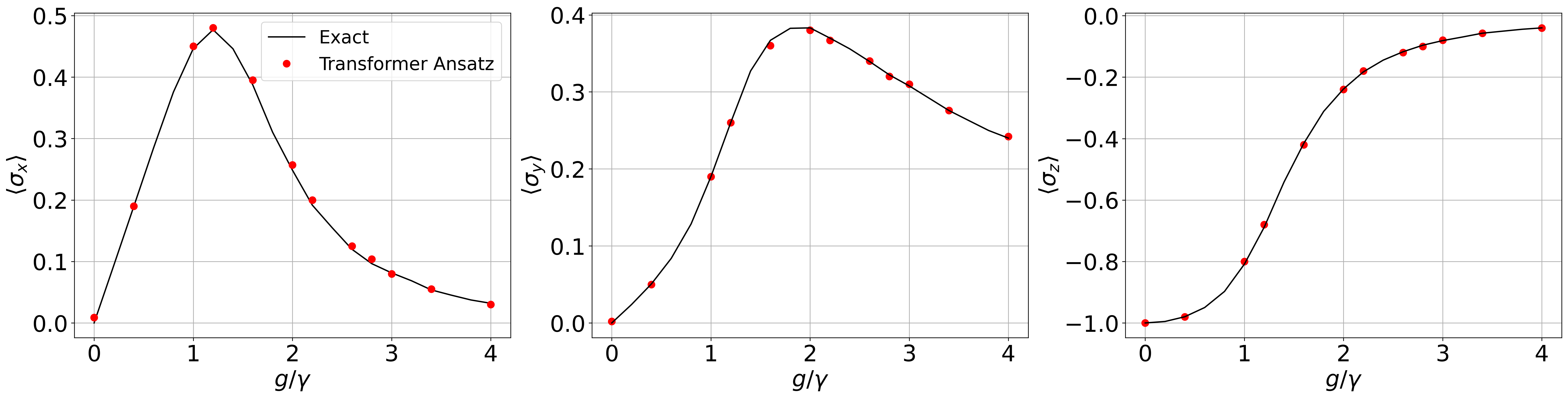}
\caption{\label{fig:sigmaz} We employ a variational transformer density operator as the steady-state ansatz for a 16-site dissipative transverse-field Ising chain with periodic boundary conditions, uniform dissipative rate, and an interaction strength of \( V = 2\gamma \). The red points in the figure represent the expectation values \(\langle \sigma_x \rangle\), \(\langle \sigma_y \rangle\), and \(\langle \sigma_z \rangle\), computed from the optimized transformer ansatz. These points precisely match the exact solid black line, demonstrating its capability to accurately capture the steady-state properties of the system.}
\end{figure*}

\vspace{1em}
\emph{Variational algorithm for searching steady state based on transformer density operator ansatz.} ---
Consider a quantum system $\mathcal{H}_S$ with $\operatorname{dim} \mathcal{H}_S = d_S$. When coupled with a Markovian environment $\mathcal{H}_E$, the evolution equation of the system takes the form of the Gorini–Kossakowski–Sudarshan–Lindblad (GKSL) equation~\cite{lindblad1976generators,gorini1976completely}, also known as the quantum Liouville equation or master equation:
\begin{equation}\label{eq:Lindblad}
    \frac{d \hat{\rho}}{d t} = \mathcal{L}(\hat{\rho}) = \frac{1}{i\hbar}[H, \hat{\rho}] + \sum_{i > 0} \gamma_i \left( L_i \hat{\rho} L_i^{\dagger} - \frac{1}{2} \{ L_i^{\dagger} L_i, \hat{\rho} \} \right),
\end{equation}
where the Lindbladian $\mathcal{L}$ is a superoperator, $H$ is the Hamiltonian, and $L_i$'s are the jump operators associated with the dissipative processes induced by the environment. The $\gamma_i$'s are the dissipation rates. There are at most $d_S^2 - 1$ jump operators over $\mathcal{H}_S$.
The GKSL equation is the most general equation satisfying the following constraints: (i) local in time, (ii) ensures the positivity $\rho(t) \geq 0$ for all $t$, (iii) is trace-preserving, i.e., $\Tr \hat{\rho}(t) = 1$ for all $t$, and (iv) forms a quantum dynamical semigroup.

In the vectorization form, we have 
\begin{equation}
    \frac{d}{dt} |\rho\rrangle=\hat{\mathcal{L}} |\rho\rrangle,
\end{equation}
where the Lindblad operator $\hat{\mathcal{L}}$ is of the form
\begin{equation}\label{eq:VecLin}
\begin{aligned}
     \hat{\mathcal{L}}=&-i(H\otimes \identity - \identity \otimes H^T)\\
    &+\sum_{i>0} \gamma_i[L_i\otimes L^*_i-\frac{1}{2}(L^{\dagger}_iL_i\otimes  \identity +  \identity \otimes L_i^TL_i^{*})].
\end{aligned}
\end{equation}
The steady state plays a crucial role in real applications and is defined as the fixed point of the dynamical semigroup, $\hat{\rho}_{SS}=\lim_{t\to \infty} \hat{\rho}(t)$. It can be equivalently expressed as the null state for the Lindbladian $\mathcal{L}$:
\begin{equation}
\hat{\mathcal{L}} \hat{\rho}_{SS} = 0.
\end{equation}
When the steady state is a pure state, it is referred to as a dark state. A dark state is decoherence-free, making it a crucial resource for quantum computing and various quantum information tasks~\cite{lidar2003decoherence, BlumeKohout2008characterizing}.

Solving for the steady state is a challenging task, especially for many-body systems in condensed matter physics. Since $\hat{\mathcal{L}}$ is generally non-Hermitian, we introduce $\mathfrak{L} = \hat{\mathcal{L}}^{\dagger} \hat{\mathcal{L}}$, which has a real and non-negative spectrum. The steady state satisfies $\mathfrak{L}|\rho_{SS}\rrangle = 0$. 
It is worth mentioning that, in general, a solution to the above equation is a state vector in the doubled Hilbert space, but it may not correspond to a valid density operator. However, for many physical systems, the uniqueness of the steady state ensures that this does not pose a significant issue \cite{Schirmer2010stablizing,Cai2013algebraic,Horstmann2013noise,prosen2012comments,hsiang2015nonequilibrium}. Using transformer density operator $\rho_{\bm{\theta}}$ as an ansatz, the loss function can be defined as
\begin{equation}
\label{eq:loss}
\operatorname{Loss}(\bm{\theta}) = \llangle \rho_{\bm{\theta}} | \mathfrak{L} | \rho_{\bm{\theta}} \rrangle.
\end{equation}
For the loss function, we have $\operatorname{Loss}(\bm{\theta}) \geq 0$, and it reaches zero if and only if the steady state is reached. The parameters $\vec{\theta}$ of the transformer that achieve this will give the desired steady state. This ansatz can be modeled by a neural network and optimized using variational Monte Carlo methods as introduced in Sec.~\ref{SM:procedure} in the supplementary material. Through variational Monte Carlo optimization, the parameters \(\bm{\theta}\) are adjusted so that the resulting density operator accurately represents the steady state of the open quantum system under study. The full optimization procedure is described in detail in Sec.~\ref{SM:OANDE} of the supplementary material.

\begin{figure}[b]
    \centering
    \includegraphics[width=0.9\linewidth]{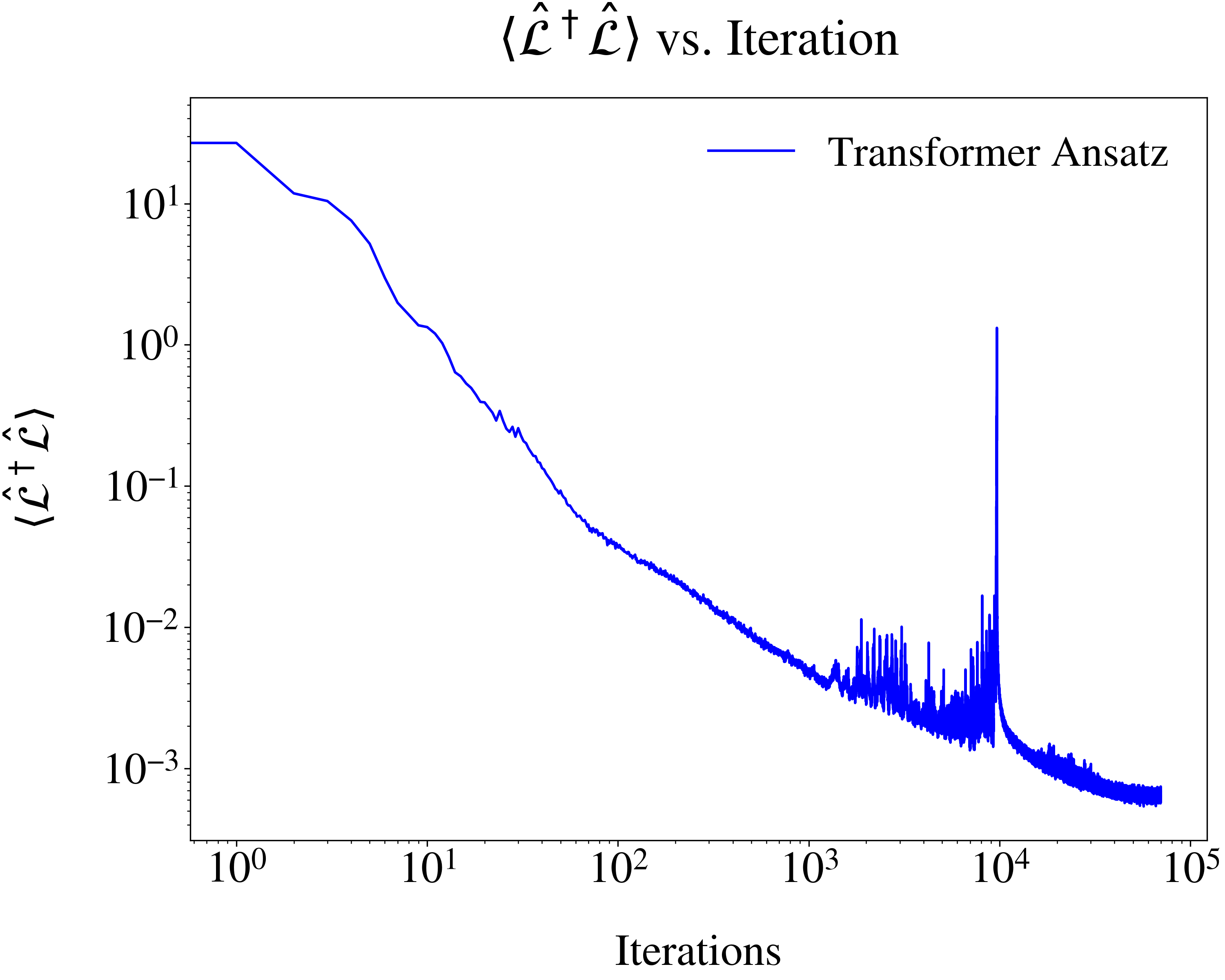} 
\caption{\label{fig:comparison1}Optimization of the variational loss function $\operatorname{Loss}(\bm{\theta}) = \langle \hat{\mathcal{L}}^\dagger \hat{\mathcal{L}} \rangle_{\rho_{\bm{\theta}}}$. We plot the optimization processes of the transformer density operator ansatz in approximating the steady-state density operator of an open quantum Ising chain. We consider a 16-site dissipative transverse-field Ising chain with periodic boundary conditions. The system has a uniform dissipation rate \(\gamma\), an interaction strength \(V = 2\gamma\), and a fixed transverse field of magnitude \(g = 1.6\). The optimization employs simple stochastic gradient descent and stochastic reconfiguration with respective fixed learning rates. As shown in the figure, although moderate fluctuations occur in the early stages of training, the loss function ultimately decreases by several orders of magnitude, demonstrating successful convergence toward the steady state.}
\end{figure}

\vspace{1em}
\emph{Numerical results for the dissipative transverse-field Ising chain.} ---
The Hamiltonian of the transverse-field Ising model is  
\begin{equation}
    H = \frac{V}{4} \sum_{i=1}^{N} \sigma_{i}^{z} \sigma_{i+1}^z + \frac{g}{2} \sum_i \sigma_i^x,
\end{equation}
where $V$ is the interaction strength, $g$ is the transverse field strength, and $\sigma_i^z$, $\sigma_i^x$ are Pauli matrices acting on the \(i\)-th spin while acting as the identity operator on all other spins in the system.  
Dissipation is introduced via local spin decay, modeled by the jump operators $L_i = \sigma^-_i = \frac{1}{2}(\sigma_i^x - i\sigma^y_i)$, where $\sigma^-_i$ is the lowering operator acting on site \(i\). The system's evolution is governed by the Lindblad master equation, with the corresponding Lindblad superoperator given by Eq.~\eqref{eq:VecLin}.

\begin{figure}
\centering
\includegraphics[width=1\linewidth]{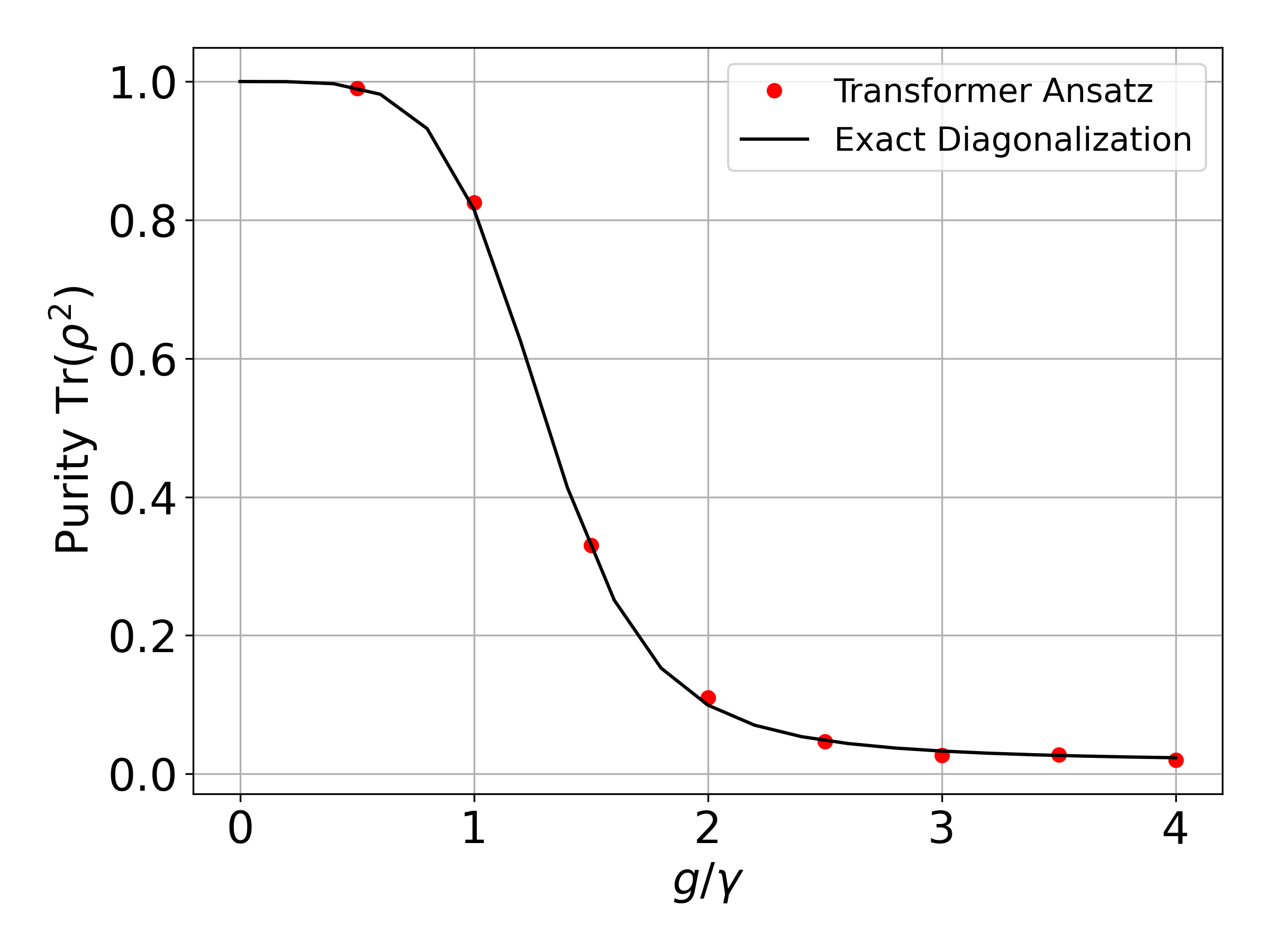}
\caption{\rev{Purity of the steady state for a 6-site transverse-field Ising chain with periodic boundary conditions, uniform local dissipative rate $\gamma$ and nearest-neighbour coupling $V=2\gamma$. The graph shows that the purity of the steady state decreases with the increase of the transverse magnetic fields and the purity from our transformer ansatz aligns well with the exact curve. This shows that our transformer ansatz has the capability of learning mixed steady states.}}
\label{fig:purity1}
\end{figure}

In our numerical simulations, we study a system of 16 lattice sites with periodic boundary conditions. The dissipation rate \(\gamma_i\) in Eq.~\eqref{eq:VecLin} is taken to be uniform across all sites, and we set the coupling constant to \(V = 2\gamma\). To obtain the steady-state density matrix, we employ our transformer density operator ansatz, optimizing its parameters using the Stochastic Gradient Descent or Adam~\cite{kingma2014adam} optimizer in combination with the stochastic reconfiguration algorithm~\cite{chen2024empowering}.

\rev{

Steady states can be either pure or mixed. To verify that our method accurately captures mixed steady states, we computed the purity $\mathrm{Tr}(\rho_{\vec{\theta}}^2)$ of the optimized transformer-based density operator ansatz, and compared it with the results obtained via exact diagonalization. The purity, which ranges from 0 to 1, quantifies the degree of mixedness: a smaller purity indicates a more mixed state.

When the transverse field is zero ($g=0$), it is straightforward to verify that the pure state $\chi_{\rm SS} = |1 \cdots 1\rangle \langle 1 \cdots 1|$ is a steady state with purity one. This follows since $[H, \chi_{\rm SS}] = 0$ and $\sigma^- |1\rangle = 0$.
As $g$ increases, the steady state gradually becomes mixed due to the system undergoing local energy dissipation. In the limit $g \gg V$, the Hamiltonian reduces to $H = \frac{g}{2} \sum_i \sigma^x_i$, in which case the qubits decouple and evolve independently. Consequently, the steady state of the full system can be obtained as a tensor product of single-qubit steady states (due to the symmetry).

The steady state condition for a single qubit reads
\begin{equation}
    -i\frac{g}{2} [\sigma^x, \rho_{\rm SS}^{(1)}] + \gamma \left( \sigma^- \rho_{\rm SS}^{(1)} \sigma^+ - \frac{1}{2} \{ \sigma^+ \sigma^-, \rho_{\rm SS}^{(1)} \} \right) = 0.
\end{equation}
For simplicity, we set $\gamma = 1$. Solving this equation yields
\begin{equation}
    \rho_{\rm SS}^{(1)} = \begin{pmatrix}
        \frac{g^2}{1 + 2g^2} & \frac{-ig}{1 + 2g^2} \\
        \frac{ig}{1 + 2g^2} & \frac{1+g^2}{1 + 2g^2}
    \end{pmatrix}.
\end{equation}
Therefore, the steady state of the full Ising chain in this limit reads
\begin{equation}
    \rho_{\rm SS} = \bigotimes_{i=1}^N \rho_{\rm SS}^{(1)}.
\end{equation}
As $g \to \infty$, the steady state approaches the maximally mixed state: $\rho_{\rm SS} \to \mathds{I}/2^N$. In this case, the purity of the state is given by
\begin{equation}
    \mathrm{Tr}(\rho_{\rm SS}^2) = \frac{1}{2^N}.
\end{equation}
For a long chain with length $N \gg 1$, the purity decays exponentially and tends to zero, indicating a highly mixed steady state.

We examined a 6-site transverse-field Ising model to demonstrate this capability. As shown in Fig.~\ref{fig:purity1}, the purity decreases monotonically as the transverse field strength $g$ increases. The solid black curve is calculated via exact diagnolization using qutip~\cite{lambert2024qutip, johansson2012qutip} and the red points are the purity calculated from our trained transformer density operator ansatz. At $g/\gamma = 4$, the purity drops to approximately 0.01, indicating a highly mixed steady state. This substantial reduction in purity confirms that the steady state is indeed mixed rather than pure, with the degree of mixing increasing with stronger transverse fields. Importantly, our transformer density operator ansatz accurately reproduces the exact purity values across the entire parameter range, demonstrating excellent agreement with the benchmark calculations. 

}

\begin{figure*}[th]
\centering
\includegraphics[width=1\textwidth]{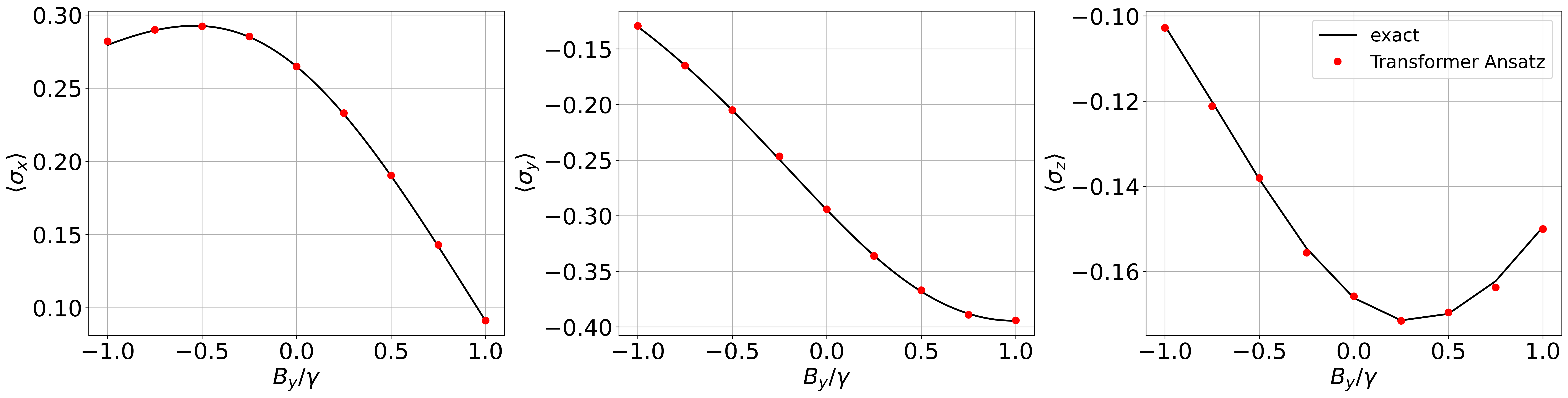}
\caption{
Expectation values of the averaged megnetization components $\langle \sigma_x \rangle$, $\langle \sigma_y \rangle$, and $\langle \sigma_z \rangle$ are evaluated as functions of the transverse magnetic field $B_y / \gamma$ for a 5-site Heisenberg spin chain, with parameters $J_x / \gamma = 1.4$, $J_y / \gamma = 2.0$, $J_z / \gamma = 1.0$, $B_x / \gamma = -1.0$, $B_z / \gamma = 0.1$ and periodic boundary conditions. Our transformer density operator ansatz (red dots) achieves excellent agreement across the entire range of $B_y / \gamma$, validating its effectiveness in approximating the quantum state and observable dynamics. 
}
\label{fig:heisen}
\end{figure*}

To further evaluate the accuracy and expressiveness of our ansatz, we compute the expectation values of local observables, specifically the steady-state magnetization components along the \(x\), \(y\), and \(z\) directions, given by $\langle {\sigma}_k \rangle_{\mathrm{ss}} = \frac{1}{N} \sum_{i=1}^{N} \langle \sigma^k_i \rangle, k \in \{x, y, z\}$, for different magnetic field strengths $g$. These expectation values are estimated via Monte Carlo sampling, following the procedure outlined in Sec.~\ref{SM:expectation} of the supplementary material. Specifically, we obtain matrix elements of the density operator for a set of sampled configurations, from which we approximate the expectation values using a local estimator approach. A detailed formulation of the method, including the probability distribution used in the sampling process and the construction of the local estimator, can be found in Secs.~\ref{SM:sample} and \ref{SM:expectation} of the supplementary material.

In Fig.~\ref{fig:sigmaz}, we present \(\langle \sigma_x \rangle\), \(\langle \sigma_y \rangle\), and \(\langle \sigma_z \rangle\) as a function of the normalized transverse field strength \(g/\gamma\). The solid black curves represent numerically exact steady-state measurement values obtained by combining the quantum-trajectory algorithm with the MPO method~\cite{vicentini2019variational, vicentini2019stochastic}. The red points show the predictions of our variational transformer density operator ansatz. Across the entire field range the measurements from our trained ansatz match closely with the black exact curve. The time complexity of the quantum-trajectory method is exponential to the system size, while the computational load of our variantional transformer ansatz grows slowly with additional spins. Therefore our ansatz effectively captures the essential physics of the steady state in the dissipative quantum system. 

To provide a more comprehensive demonstration of the performance of our ansatz, we also show the optimization curve for a transverse field of magnitude \(g = 1.6\), as presented in Fig.~\ref{fig:comparison1}, which illustrates the convergence of the loss function $\operatorname{Loss}(\bm{\theta}) = \langle \hat{\mathcal{L}}^\dagger \hat{\mathcal{L}} \rangle_{\rho_{\bm{\theta}}}$ as a function of the number of iterations. Detailed optimization strategy is introduced in Sec.~\ref{SM:sche} of the supplementary material. The transformer density operator ansatz of the blue curve demonstrates strict convergence. This indicates that the transformer density operator ansatz is well equipped to capture correlations inherent in the quantum Ising chain, owing to its self-attention mechanism. Such correlations are essential for accurately modeling the steady-state properties of dissipative quantum systems. This makes the transformer density operator a promising ansatz for addressing systems with complex correlation structures.

\vspace{1em}
\emph{Numerical results for dissipative Heisenberg spin chain model.} ---
We then test our model on the Heisenberg lattice spin system. The system under consideration is governed by the following Hamiltonian:
\begin{equation}
H = \sum_{i=1}^N \sum_{k=x,y,z} \left( J_k \sigma_i^k \sigma_{i+1}^k + B_k \sigma_i^k \right),
\end{equation}
where \(J_k\) denotes the interaction strength for the spin component along the \(k\)-th axis (\(k = x, y, z\)) between nearest-neighbor spins \(i\) and \(i+1\). The term \(B_k \sigma_i^k\) represents the effect of an external magnetic field applied along the \(k\)-th direction at site \(i\), with \(B_k\) being the corresponding field strength. This Hamiltonian captures both the anisotropic exchange interactions and the influence of an external magnetic field on the quantum spin system. For the dissipative part, we consider a uniform dissipation rate across all sites, setting \(\gamma_j = \gamma\) for all \(j = 1, \dots, N\). The corresponding jump operators are chosen as \(L_j = \sigma_j^-\), representing local spin lowering at each site.

In our numerical test on the Heisenberg lattice spin system, we evaluate the observables $\langle \sigma_x \rangle$, $\langle \sigma_y \rangle$, and $\langle \sigma_z \rangle$ as functions of the transverse magnetic field ratio $B_y / \gamma$. The system consists of $N=5$ sites, with interaction strengths $J_x / \gamma = 1.4$, $J_y / \gamma = 2.0$, and $J_z / \gamma = 1.0$. The other components of the local magnetic field vector are set to $B_x / \gamma = -1.0$ and $B_z / \gamma = 0.1$. We employ the same model structure as in the previous example of the Ising model. For optimization, we use the Adam optimizer in combination with the stochastic reconfiguration method~\cite{chen2024empowering} without a scheduler, which is enough to ensure stable convergence during the variational minimization process.

The experimental results, presented in Fig.~\ref{fig:heisen}, compare the performance of our proposed transformer density operator ansatz with the exact baseline. The exact curve (black line) is obtained by the iterative BiCGStab solver of NetKet as described in \ref{SM:Ben} in the supplementary material. This method can give results with high accuracy for system of size larger than 7 and we utilize it as our baseline. As shown in the graph, the result from our ansatz matches the exact curve well. In the supplementary material~\ref{sup:more}, we also provided the result of 10-site one-dimensional Heisenberg model. These results highlight the robustness of the transformer density operator ansatz in capturing the complex features of the spin system across varying magnetic field strengths. 

\rev{
\vspace{1em}
\emph{Numerical results for 2D dissipative Ising model.} ---
To demonstrate the scalability of our approach to higher-dimensional systems, we extend our study to a 2D dissipative Ising model defined on an $M \times N$ square lattice. The Hamiltonian takes the form
\begin{equation}
    H = \frac{V}{4} \sum_{\langle i j \rangle} \sigma^z_i \sigma^z_j + \frac{g}{2} \sum_{i \in \mathbb{Z}_M \times \mathbb{Z}_N} \sigma^x_i,
\end{equation}
where the notation $\langle i j \rangle$ indicates a summation over all nearest-neighbor pairs. 
The local jump operator is taken to be $L_i = \sigma^-_i$, as discussed in, e.g., Ref.~\cite{Roberts2023exact}.

We investigate a $2 \times 3$ lattice system with periodic boundary conditions. The system parameters are set to $V = 2\gamma$ and we vary the transverse field strength $g/\gamma$ to explore different regimes of the phase diagram. For this system size, exact diagonalization becomes computationally demanding but remains feasible for benchmark comparison.

\begin{figure*}[th]
\centering
\includegraphics[width=1\textwidth]{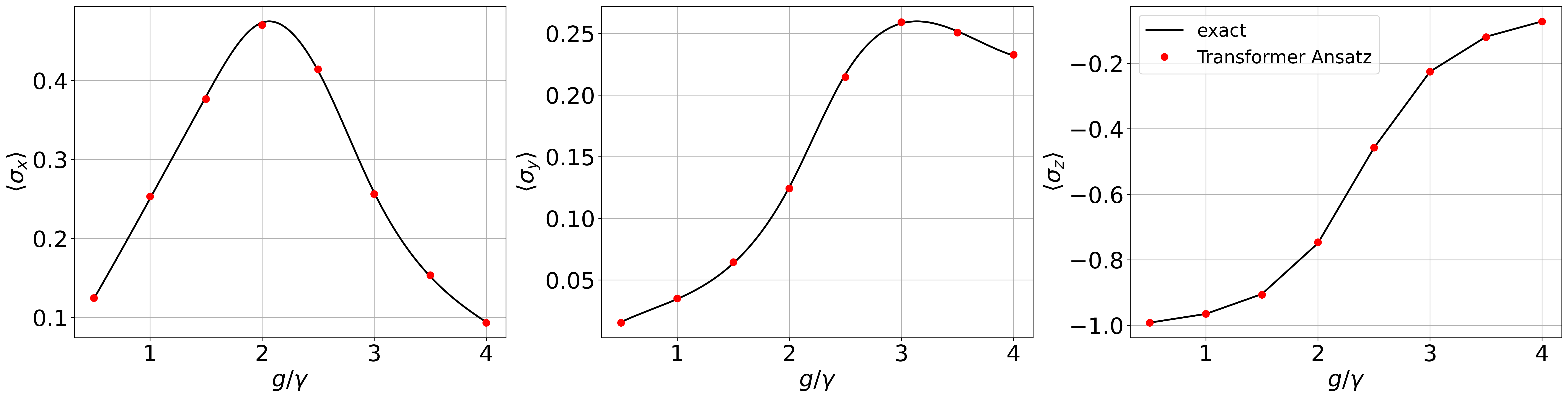}
\caption{\label{fig:2d_ising} \rev{Expectation values of the averaged magnetization components $\langle \sigma_x \rangle$, $\langle \sigma_y \rangle$, and $\langle \sigma_z \rangle$ as functions of the normalized transverse field strength $g/\gamma$ for a 2D dissipative Ising model on a $2 \times 3$ square lattice with periodic boundary conditions. The system parameters are $V = 2\gamma$ with uniform dissipation rate $\gamma$ across all 6 sites. The red dots represent results from our transformer density operator ansatz. The transformer architecture successfully handles the increased complexity of 2D geometry, including enhanced coordination numbers and frustration effects compared to 1D systems.}
}
\end{figure*}

As shown in Fig.~\ref{fig:2d_ising}, our transformer density operator ansatz demonstrates excellent accuracy in capturing the steady-state magnetization components $\langle \sigma_x \rangle$, $\langle \sigma_y \rangle$, and $\langle \sigma_z \rangle$ across the entire range of transverse field strengths. The exact curve is calculated via the iterative BiCGStab solver of NetKet and qutip. The 2D geometry introduces additional complexity through the increased coordination number (four nearest neighbors per site) and enhanced frustration effects, yet our transformer approach handles these challenges effectively. We also include the result for $2 \times 2$ in the supplementary material~\ref{sup:more}. For large 2D system, more parameters are required to reach a high precision.

The self-attention mechanism proves particularly advantageous in the 2D setting, as it naturally captures the long-range correlations that emerge from the interplay between the 2D geometry and dissipative dynamics.

}

\vspace{1em}
\emph{Conclusion and discussion.} ---
In this work, we introduce a transformer variational ansatz to efficiently encode and solve the steady states of dissipative quantum systems. By vectorizing the density operator into a many-body state, we embed the input configurations into feature vectors and then utilize the self-attention mechanisms to model long-range correlation, which is crucial in the open system. Numerical experiments on paradigmatic models, including one-dimensional dissipative transverse-field Ising and Heisenberg models, as well as a two-dimensional dissipative Ising model, show that our approach accurately reproduces the mixed steady states of various dissipative systems while maintaining a compact parameterization and substantial complexity advantage.

Several promising avenues for future research exist. First, extending the current scheme to systems with more complex interactions, such as long-range couplings or higher-dimensional lattices, could uncover richer dynamical and correlation structures. 
Second, exploring systems with more intricate boundary conditions (especailly in two and higher dimensions) would provide further insights into the robustness and expressiveness of the transformer density operator ansatz. 
Third, adapting this framework to predict unknown quantum states by refining the cost function could offer a novel approach to quantum state reconstruction and tomography. Finally, incorporating advanced techniques like self-supervised learning, hyperparameter optimization, or attention-based modules tailored to specific physical symmetries may enhance the model's generalization capability. We expect that this flexible framework will serve as a strong foundation for tackling more complex open quantum systems and advancing our understanding of dissipative many-body physics.

\vspace{1em}
\emph{Acknowledgments.} ---
We acknowledge Di Luo, Filippo Vicentini, Yuan-Hang Zhang,  and Chen Zhuo for beneficial communications. We thank Filippo Vicentini and Di Luo for sharing their codes with us. The numerical implementation of the variational transformer density operator ansatz was done using JAX. The variational quantum Monte Carlo and stochastic reconfiguration optimizers are available in NetKet.
Z. J. and D. K. are supported by the National Research Foundation in Singapore and A*STAR under its CQT Bridging Grant, CQT- Return of PIs EOM YR1-10 Funding and  CQT Young Researcher Career Development Grant.

\appendix

\section{Steady state and the vectorization of density operator}
\label{sec:steadyLind}

In this section, we review the vectorization formalism of the density operator and its dynamics, a powerful framework for representing the steady states of dissipative open quantum systems. This approach has recently gained significant attention in studies of open-system quantum phases, weak and strong symmetries, tenfold classification, and other related topics (see, e.g., \cite{Kawabata2023symmetry,bao2023mixedstate,Sohal2025noisy,Yoshioka2019neural,ma2024SPTmixed,nigro2020complexity,nigro2019uniqueness}). Unlike the traditional density matrix representation, the vectorized form offers a more convenient computational framework for steady-state analysis. By expressing the density operator in this form, steady states can be obtained by solving for the ground state of a specially constructed operator. 

To solve for the steady state $\hat{\rho}_{SS}$ of a Lindbliadian $\mathcal{L}$, we introduce the vectorization of the density operator $\hat{\rho}$ in a fixed basis $\{|\alpha\rangle\}$. Given the representation  
\begin{equation}
\hat{\rho} = \sum_{\alpha,\beta} \rho_{\alpha,\beta} |\alpha\rangle \langle \beta|,
\end{equation}
The vectorized form is defined as  
\begin{equation}
|\rho\rrangle = \sum_{\alpha,\beta} \rho_{\alpha,\beta} |\alpha\rangle |\beta\rangle.
\end{equation}
More generally, we have  
\begin{equation}
|(|\psi\rangle\langle \phi|)\rrangle = |\psi\rangle |\phi^*\rangle,
\end{equation}
where $\phi^*$ is the complex conjugate of $\phi$ in the given basis. It is clear that vectorization is a basis-dependent operation.

Let \( A \) and \( B \) be two operators acting on separate subsystems. Their vectorized forms are denoted as \( |A\rrangle \) and \( |B\rrangle \), respectively. Note that in the vectorization process, the reordering of kets and bras for each local degree of freedom must be taken into account. Consequently, the vectorization of the tensor product does not satisfy a simple factorization
\begin{equation}
    |A \otimes B\rrangle \neq |A\rrangle \otimes |B\rrangle.
\end{equation}
This distinction arises because vectorization is performed in a specific basis, and care must be taken in handling the ordering of indices when working with composite systems.

In the vectorized form, a superoperator acting as $A \rho B$ is represented as  
\begin{equation}
    A \rho B \mapsto (A \otimes B^T) |\rho\rrangle.
\end{equation}
For a multipartite system of $N$ spins, the vectorized representation of the density operator is given by  
\begin{equation}
|\rho \rrangle = \sum_{\alpha_1,\dots,\alpha_N; \beta_1,\dots,\beta_N} \rho_{\alpha_1,\dots,\alpha_N;\beta_1,\dots,\beta_N} |\alpha_1,\dots,\alpha_N\rangle \otimes |\beta_1,\dots,\beta_N \rangle.
\label{eq:KFRapp}
\end{equation}
The vectorized form encodes the state of multiple subsystems by introducing auxiliary degrees of freedom.

Consider the Gorini–Kossakowski–Sudarshan–Lindblad (GKSL) equation (set $\hbar=1$)
\begin{equation}
    \frac{d \rho}{d t}=\mathcal{L}(\rho)={-i}[H, \rho]+\sum_{i>0}\gamma_i( L_{i} \rho L_{i}^{\dagger}-\frac{1}{2}\{L_{i}^{\dagger} L_{i}, \rho\}),
\end{equation}
In the vectorization form, we have 
\begin{equation}
    \frac{d}{dt} |\rho\rrangle=\hat{\mathcal{L}} |\rho\rrangle,
\end{equation}
where the Lindblad operator $\hat{\mathcal{L}}$ is of the form
\begin{equation}
\begin{aligned}
     \hat{\mathcal{L}}=&-i(H\otimes \identity -\identity \otimes H^T)\\
   & +\sum_{i>0} \gamma_i[L_i\otimes L^*_i-\frac{1}{2}(L^{\dagger}_iL_i\otimes \identity +\identity \otimes L_i^TL_i^{*})].
\end{aligned}
\end{equation}
The density matrix in the doubled Hilbert space is the state that
\begin{equation}\label{eq:ori}
    \hat{\mathcal{L}}|\rho \rrangle =0.
\end{equation}
It is important to highlight that, in general, a solution to the above equation corresponds to a state vector in the doubled Hilbert space, but it may not always be a valid density operator. However, steady states of open systems may not always be unique. The uniqueness of \(|\rho_{\text{ss}}\rrangle\) depends on the spectral properties of \(\hat{\mathcal{L}}\), and certain symmetries can lead to degenerate steady states. In cases where the steady state is not unique, additional selection rules or symmetry constraints may be necessary to determine the correct physical solution. Many open systems have a unique steady state \cite{Schirmer2010stablizing,Cai2013algebraic,Horstmann2013noise,prosen2012comments,hsiang2015nonequilibrium}, which guarantees that the resulting state in the doubled Hilbert space automatically corresponds to a steady-state density operator.

Since \(\hat{\mathcal{L}}\) is generally non-Hermitian, its eigenvalues are complex. Directly solving \(\hat{\mathcal{L}} |\rho\rrangle = 0\) may lead to numerical instability. Instead, we introduce the Hermitian operator
\begin{equation}
   \mathfrak{L} =  \hat{\mathcal{L}}^{\dagger}\hat{\mathcal{L}},
\end{equation}
where $\mathfrak{L}$ is a Hermitian matrix with real and non-negative eigenvalues. The zero-eigenvalue solution of \(\mathfrak{L} |\rho \rrangle = 0\) corresponds to the steady-state solution of the original Lindblad equation. This formulation enables the use of variational minimization techniques to efficiently approximate the steady state.

The lowest eigenstate with eigenvalue $\lambda = 0$ of $\hat{\mathcal{L}}^{\dagger}\hat{\mathcal{L}}$ corresponds to the steady state. Therefore, solving the equation
\begin{equation}
\hat{\mathcal{L}}^{\dagger}\hat{\mathcal{L}}|\rho \rrangle = 0
\end{equation}
provides the steady state, as shown in equation \eqref{eq:ori}. We can then optimize the energy functional to find the ground state, akin to the approach in closed systems. The energy functional is given by
\begin{equation}
    E = \llangle \rho |\mathfrak{L}|\rho \rrangle = \llangle \rho |\hat{\mathcal{L}}^{\dagger}\hat{\mathcal{L}} |\rho \rrangle.
\end{equation}
This expression serves as the optimization objective and loss function, represented as the expectation value of the operator $\hat{\mathcal{L}}^{\dagger}\hat{\mathcal{L}}$ in the vectorized form.

In the vectorized form, the expectation value of an observable is calculated as
\begin{equation}
    \Tr (A^{\dagger}B) = \llangle A | B \rrangle.
\end{equation}
The expectation value of an observable \(O\) in the vectorized formalism is given by
\begin{equation}
    \langle O \rangle =\Tr (O\rho) = \llangle O |\rho \rrangle,
\end{equation}
where the ``$\dagger$" has been omitted for the second equality since $O$ is an Hermitian operator.

\section{Transformer Density Operator Architecture}
\label{sec:TansformerDOA}

In this work, we employ a transformer density operator approach to parameterize the steady state of spin chains with periodic boundary conditions. While convolutional layers primarily capture local dependencies through learnable filters, the \emph{multi-head self-attention} modules—adapted from the \emph{transformer} framework—enable the model to capture dependencies across the entire system. This makes them particularly effective for representing quantum states with complex interactions. By dynamically extracting multiple similarity patterns along the chain, these modules can model both short- and long-range correlations. This capability is especially crucial in systems with periodic boundary conditions, where distant spins remain strongly correlated, necessitating a global perspective for accurate representation.

Concretely, the network begins by embedding each spin configuration pair $(\sigma, \sigma')$ into a continuous feature space using a shallow CNN stage. The resulting features are then passed through transformer-based attention layers, which aggregate information across all sites in a translation-invariant manner. This attention mechanism naturally captures various forms of spin correlation, as each attention head can learn to focus on different regions or subsets of sites within the chain. Finally, a small dense module outputs the complex amplitudes (or the real and imaginary components) that define the desired quantum state or density operator. This design combines the local feature extraction capabilities of CNNs with the global context modeling power of multi-head self-attention, making the network particularly well-suited for representing the steady-state properties of open quantum spin systems.

\subsubsection{Overview of the Transformer Density Operator Architecture}

In the standard transformer \cite{vaswani2023attentionneed}, positional encodings, multi-layer encoder-decoder blocks, and feed-forward networks are typically used to capture long-range correlations in sequential data. However, \emph{positional encoding} and the \emph{decoder} mechanism are not used for steady-state representations. We modify the transformer architecture to respect the symmetries of the steady-state density operator while preserving its ability to capture long-range correlations.

Specifically, we introduce a transformer density operator that replaces conventional positional encoding with a translation-invariant representation and focuses on self-attention mechanisms to learn multiple similarity patterns across spins. The network first applies two convolutional blocks to encode input qubit configurations while respecting periodic boundary conditions. A self-attention block then captures rich correlation patterns across the entire spin chain. Next, a global mean-pooling step enforces translation invariance. Finally, a dense layer maps the extracted features to produce the real and imaginary components of the complex amplitudes. To ensure Hermitian symmetry in the steady-state density operator, the network computes the logarithm of the sum of two symmetrized exponentials of these amplitudes to obtain an ansatz density operator.

\subsubsection{Feature Embedding}
To encode $\text{B}$ batches of configurations of a vectorized steady state \((\mathbf{x}^{(\ell)},\mathbf{x}^{(r)})\in\mathbb{R}^{B\times 2L}\) of a one-dimensional chain with \(L\) spins, we stack these two spin configurations \(\mathbf{x}^{(\ell)},\mathbf{x}^{(r)}\in\mathbb{R}^{B\times L}\) as a two-channel input
\[
(\mathbf{x}^{(\ell)}, \mathbf{x}^{(r)}) \quad\longmapsto\quad 
\underbrace{\left[\mathbf{x}^{(\ell)}_1, \ldots, \mathbf{x}^{(\ell)}_L\right]}_{\text{channel 1}}
\;\vert\;
\underbrace{\left[\mathbf{x}^{(r)}_1, \ldots, \mathbf{x}^{(r)}_L\right]}_{\text{channel 2}},
\;\in\;\mathbb{R}^{B \times L \times 2}\]
effectively producing a \(\;L\times 2\; \) array. 
This two-dimensional format allows convolutional layers to scan each pair \((\mathbf{x}^{(\ell)}_i,\mathbf{x}^{(r)}_i)\) jointly. A dummy dimension was added.

We then apply circular padding in the $L$ dimension to respect the periodic boundary conditions. Specifically,
\begin{equation}\label{eq:circ-conv}
\mathbf{X}^{(1)} \;=\;\mathrm{Conv}_{\text{circular}}\!\bigl(\mathbf{X}_{\mathrm{input}}\bigr)
\;\in\;\mathbb{R}^{B\times L\times 1\times C_1},
\end{equation}
where $C_1$ is the number of filters, followed by a non-linear activation. 
A second convolution with $C_2$ filters is applied in the same manner:
\begin{equation}
\mathbf{X}^{(2)} \;=\;\mathrm{Conv}_{\text{circular}}\!\bigl(\mathbf{X}^{(1)}\bigr)
\;\in\;\mathbb{R}^{B\times L\times 1\times C_2}.
\end{equation}
These layers extract local patterns by sliding kernels of size \(\,(2\times 1)\) across the stacked spin inputs. Hence, short-range correlations are encoded in a hierarchy of convolutional feature maps.

\subsubsection{Self-Attention for Global Correlations}
In open quantum systems with periodic boundary conditions, the system may exhibit different scales of correlation due to competition between spin-spin interactions, external fields, and dissipative channels.
To capture the variety of correlation patterns among spins, we incorporate a self-attention module that calculates a dot-product attention among the spin-site embeddings. In particular, we use either a \emph{single-head} attention layer or a \emph{multi-head} architecture that splits the hidden dimension into several heads, each learning a distinct similarity pattern.  
This allows any spin site to directly attend to all others, thus modeling extended or global correlations 
that are often found in the steady state of the dissipative spin chain. 
We reshape \(\mathbf{X}^{(2)}\) by removing the dummy axis:
\begin{equation}
\mathbf{X}\;=\;\mathrm{squeeze}\!\bigl(\mathbf{X}^{(2)},\,\texttt{axis}=2\bigr)
\;\in\;\mathbb{R}^{B \times L \times C_2}.
\end{equation}
We then apply either a single-head or multi-head self-attention block. 
Let \(\mathbf{Q},\mathbf{K},\mathbf{V}\in\mathbb{R}^{B\times L\times(C_2/h)}\) be the query, key, and value embeddings for $h$ heads, obtained via learned linear projections:
\begin{align}
\mathbf{Q} &= \mathbf{X} \,\mathbf{W}^Q, \quad
\mathbf{K} \;=\; \mathbf{X}\,\mathbf{W}^K,\quad
\mathbf{V} \;=\; \mathbf{X}\,\mathbf{W}^V,
\end{align}
where $\mathbf{W}^Q,\mathbf{W}^K,\mathbf{W}^V \in\mathbb{R}^{C_2\times(C_2/h)}$ in each attention head.
For each head $i$, the attention weights
\begin{equation}\label{eq:attn-head-appendix}
\mathrm{head}_i \;=\;
\mathrm{softmax}\!\Bigl(\frac{\mathbf{Q}_i\,\mathbf{K}_i^\top}{\sqrt{C_2/h}}\Bigr)\,\mathbf{V}_i
\end{equation}
are computed, then concatenated and projected back to dimension $C_2$, with a final residual connection:
\begin{equation}
\mathbf{X}' \;=\;\mathrm{Concat}\big(\mathrm{head}_1,\dots,\mathrm{head}_h\big)\,\mathbf{W}^O \;+\;\mathbf{X}.
\end{equation}
Here, \(\mathbf{W}^O\) acts as a learned linear transformation to map the concatenated multi-head outputs back to the original embedding dimension. The resulting tensor \(\mathbf{X}'\) encodes long-range correlations between all sites, and we omit additional feed-forward sub-layers and normalization for simplicity.

\subsubsection{Global Mean-Pooling and Final Output}
Although the convolutional filters reuse parameters across different lattice sites, the feature maps still encode a positional footprint. 
To impose strict \emph{translation invariance}, we add a global average pooling operation over the spatial dimension
\begin{equation}\label{eq:global-pool-final}
\mathbf{X}^{(\mathrm{pool})}
\;=\;\frac{1}{L}\sum_{j=1}^L
\bigl(\mathbf{X}'_{:\,,j\,,:\,}\bigr),
\quad
\mathbf{X}^{(\mathrm{pool})} \,\in\,\mathbb{R}^{B\times C_2}.
\end{equation}
Consequently, the final output of the transformer density operator becomes independent of site indexing. 
This design not only enforces physical symmetry under cyclic shifts but also reduces the fully connected layer’s parameters 
and enables \emph{transfer learning} to systems of different sizes \cite{mellak2024deep}. After pooling, we flatten the feature maps and feed them into a dense layer with two output neurons, 
\(
[F_0,\,F_1]
\),
representing real and imaginary components of a complex number, which is then used to generate the final complex amplitude $\rho_{\bm{\theta}}$ of the steady state. 
Hence, we obtained a complex number
\(
z
=
F_0 + i\,F_1
\) after the global pooling.
In this way all parameters remain real-valued, which simplifies optimization routines. However, in our steady-state representation, we further combine these amplitudes via 
\(
\log\!\bigl(\exp(z_1)+\exp(z_2)^{*}\bigr),
\)
where \(z_1,z_2\) are the complex outputs from \(\mathbf{x}^{(\ell)}\) and \(\mathbf{x}^{(r)}\) of a different order. 
This construction naturally enforces Hermiticity and is well-suited to describing the density operator of an open-system quantum spin chain.

\section{Optimization and Evaluation of Transformer Density Operator Ansatz}

\label{SM:OANDE}
In this section, we present the full procedure for training and validating the transformer density operator ansatz. We begin by outlining our Metropolis-Hastings sampling strategy for mixed states, which enables efficient estimates of both expectation values and gradients. We then explain how to compute observables in the mixed-state setting and describe our use of stochastic reconfiguration (also known as natural gradient descent) to stabilize and accelerate optimization. Finally, we discuss two benchmark approaches—\emph{Exact Diagonalization} and the \emph{iterative BiCGStab method}—against which we compare our results to confirm the accuracy and scalability of our approach.

\subsection{Efficient Sampling Strategy}
\label{SM:sample}
Instead of the autoregressive sampling method~\cite{Zhang2023Transformer}, the Markov chain Monte Carlo approach based on the Metropolis-Hastings algorithm, implemented via NetKet's \texttt{MetropolisLocal} sampler, is employed here. This method generates configurations by proposing local updates to the spin configuration and accepting them according to the Metropolis rule. These configurations are then used to estimate expectation values and gradients afterwards.

Given the variational density operator ansatz with current parameters $\vec{\theta}$:
\begin{equation}
    |\rho_{\vec{\theta}}\rrangle = \sum_{\valpha,\vbeta} \rho_{\vec{\theta}}(\valpha, \vbeta) |\valpha\rangle \otimes |\vbeta\rangle.
\end{equation}
We sample from the probability distribution given by:
\begin{equation}
    P_{\vec{\theta}}(\valpha, \vbeta) \propto |\rho_{\vec{\theta}}(\valpha, \vbeta)|^2
\end{equation}
The Metropolis-Hastings algorithm generates a sequence of configurations according to $P_{\vec{\theta}}(\valpha, \vbeta)$ by iteratively proposing and accepting new configurations. The algorithm works in the following steps:
\begin{enumerate}
    \item A local spin configuration update $(\valpha, \vbeta) \to (\valpha', \vbeta')$ is proposed.
    \item The new configuration is accepted with probability:
    \begin{equation}
    \begin{aligned}
        & \mathbf{A}_{\textbf{accept}}((\valpha, \vbeta) \to (\valpha', \vbeta')) \\
         =& \min \left( 1,~ \frac{P_{\vec{\theta}}(\valpha', \vbeta') g((\valpha, \vbeta) | (\valpha', \vbeta'))}{P_{\vec{\theta}}(\valpha, \vbeta) g((\valpha', \vbeta') | (\valpha, \vbeta))} \right),
    \end{aligned}
    \end{equation}
    where $g((\valpha', \vbeta') | (\valpha, \vbeta))$ is the probability of proposing $(\valpha', \vbeta')$ given $(\valpha, \vbeta)$. Since the algorithm only modifies a single spin degree of freedom per step, this transition kernel is symmetric, simplifying the acceptance ratio.
    \item If accepted, the configuration is updated; otherwise, the previous state is retained.
    \item This process is repeated to generate a Markov chain of configurations for estimating expectation values.
\end{enumerate}

This sampling strategy efficiently explores the configuration space of the variational density operator ansatz.
\subsection{Mixed-State Observables}
\label{SM:expectation}
When evaluating observables for a mixed-state density operator, one can exploit a slightly different identity that rewrites the quantum expectation value as a classical expectation over the distribution given by the diagonal of \(\hat{\rho}_{\vec{\theta}}\). Specifically, for an operator \(\hat{A}\), the expectation value can be expressed as
\begin{equation}
    \langle \hat{A} \rangle = \frac{\Tr\bigl(\hat{\rho}_{\vec{\theta}} \,\hat{A}\bigr)}{\Tr\bigl(\hat{\rho}_{\vec{\theta}}\bigr)}
    = \sum_{\valpha \in \mathbf{M}} \frac{\rho_{\vec{\theta}}(\valpha, \valpha)}{\Tr\bigl(\hat{\rho}_{\vec{\theta}}\bigr)}\,
    \tilde{A}_{\rho_{\vec{\theta}}}(\valpha),
\end{equation}
where the \emph{local estimator} is defined as
\begin{equation}
    \tilde{A}_{\rho_{\vec{\theta}}}(\valpha)
    = \sum_{\vbeta} \frac{\rho_{\vec{\theta}}(\valpha,\vbeta)}{\rho_{\vec{\theta}}(\valpha,\valpha)}
    \langle \vbeta \vert \hat{A} \vert \valpha\rangle.
\end{equation}
Here, \(\valpha\) and \(\vbeta\) label basis configurations in the Hilbert space. The probability distribution 
\[
    P_{\vec{\theta}}(\valpha)
    = \frac{\rho_{\vec{\theta}}(\valpha,\valpha)}{\Tr(\hat{\rho}_{\vec{\theta}})}
\]
plays the role of a classical distribution over the diagonal elements of \(\hat{\rho}_{\vec{\theta}}\), and the local estimator \(\tilde{A}_{\rho_{\vec{\theta}}}(\valpha)\) here involves an inner sum over all basis states (or a suitably chosen subset) to capture the off-diagonal contributions \(\rho_{\vec{\theta}}(\valpha,\vbeta)\) to the observable \(\hat{A}\). In this manner, the quantum expectation value reduces to a standard classical average over a series of sampled configurations \(\mathbf{M}\), allowing one to use the same Metropolis-Hastings sampling scheme described in Sec.~\ref{SM:sample} to estimate both observables and their gradients for the mixed-state variational ansatz.

\subsection{Metric Tensor in Stochastic Reconfiguration}
\label{SM:metric}

The stochastic reconfiguration method, also known as the natural gradient descent, introduces a metric tensor $S$ that accounts for the curvature of the variational parameter space. This tensor approximates the Fisher information matrix and ensures that the optimization follows a path that respects the geometry of the variational manifold.

The metric tensor $S$ is defined as the covariance matrix of the logarithmic derivatives of the variational wavefunction:
\begin{equation}
    S_{ij} = \langle \Delta O_i^* \Delta O_j \rangle - \langle \Delta O_i^* \rangle \langle \Delta O_j \rangle,
\end{equation}
where 
\begin{equation}
    \Delta O_i = \frac{\partial \log \rho_{\vec{\theta}}(\valpha, \vbeta)}{\partial \theta_i}
\end{equation}
is the derivative of the log-probability with respect to the variational parameters $\theta_i$.

In the context of mixed-state variational ansatz, this metric tensor is computed using Monte Carlo sampling as described in Sec.~\ref{SM:expectation}. Once $S$ is constructed, it is used in the natural gradient update step to precondition the gradient of the energy functional.

\subsection{Regularization and Stabilization}
\label{SM:SR}

To ensure numerical stability in the inversion of the metric tensor $S$, a small regularization term was introduced by adding a diagonal shift $\lambda$~\cite{sorella1998green}:
\begin{equation}
    S' = S + \lambda \identity,
\end{equation}
where $\identity$ is the identity matrix and $\lambda$ is a small positive constant. This regularization prevents the metric tensor from becoming singular and ensures robust optimization updates.

This is implemented as the stochastic reconfiguration method as a gradient preconditioner in \texttt{NetKet}, where $S$ is constructed from Monte Carlo estimates as described in Section~\ref{SM:metric}. This approach stabilizes the training dynamics of the transformer density operator ansatz and improves convergence in the steady-state optimization.

\subsection{Optimization Procedure}
\label{SM:procedure}

The training of the transformer Density Operator Ansatz is carried out in a variational framework. Our goal is to find the steady state that satisfies
\begin{equation}
    \hat{\mathcal{L}} |\rho_{\vec{\theta}}\rrangle = \hat{\mathcal{L}} \vec{\rho}_{\vec{\theta}} = 0,
\end{equation}
by minimizing the loss functional
\begin{equation}
    \operatorname{Loss}(\vec{\theta}) = \llangle \rho_{\vec{\theta}} | \mathfrak{L} | \rho_{\vec{\theta}} \rrangle,
\end{equation}
where the squared Lindblad superoperator is defined as
\begin{equation}
    \mathfrak{L} = \hat{\mathcal{L}}^{\dagger}\hat{\mathcal{L}}.
\end{equation}
with \(\hat{\mathcal{L}}\) being the Lindblad superoperator in its vectorized form. Minimizing this loss function is equivalent to minimizing the Frobenius norm of the time derivative of the density matrix (To distinguish it from the previously mentioned loss function, we refer to it here as a cost function, though both terms are interchangeable in the context of machine learning.):
\begin{equation}
\label{eq:lossnetk}
    \operatorname{Cost}(\vec{\theta}) = \frac{\|\hat{\mathcal{L}}\vec{\rho}_{\vec{\theta}}\|_2^2}{\|\vec{\rho}_{\vec{\theta}}\|_2^2} = \frac{\Tr\Bigl( \vec{\rho}_{\vec{\theta}}^{\dagger}\hat{\mathcal{L}}^{\dagger}\hat{\mathcal{L}}\vec{\rho}_{\vec{\theta}} \Bigr)}{\Tr\Bigl( \vec{\rho}_{\vec{\theta}}^{\dagger}{\vec{\rho}_{\vec{\theta}}} \Bigr)},
\end{equation}
which reaches its global minimum when the steady-state condition \(\hat{\mathcal{L}}\vec{\rho}_{\vec{\theta}}=0\) holds. 

In our implementation, we use a hybrid optimization approach that combines standard first-order gradient updates with second-order corrections via stochastic reconfiguration introduced in \ref{SM:SR}. The Stochastic Gradient Descent or Adam optimizer performs standard gradient updates, while the stochastic reconfiguration accounts for the curvature of the variational manifold by introducing a metric tensor $S$, effectively implementing a natural gradient descent strategy. Their methods are provided in NetKet via a dedicated variational driver \texttt{nk.SteadyState} and Eq.~\eqref{eq:lossnetk} is the cost function that NetKet uses in its steady-state driver.

\subsubsection{Hybrid Optimization Approach}

At each optimization step, the parameters \(\vec{\theta}\) are updated using two complementary components:

\paragraph{First-Order Gradient Update:}  
Standard gradient-based methods are used to update the parameters
\begin{equation}
    {\vec{\theta}}^{(k+1)} = {\vec{\theta}}^{(k)} - \eta \nabla_{\vec{\theta}} \operatorname{Cost}(\vec{\theta}),
\end{equation}
where \(\eta\) is the learning rate. In our experiments, we primarily use stochastic gradient descent with an appropriate learning rate schedule, while Adam is also applied in our Heisenberg model example. 

\paragraph{Second-Order Correction via Stochastic Reconfiguration:}  
To account for the geometry of the variational parameter space, we incorporate stochastic reconfiguration, which introduces a metric tensor \(S\) that is an approximation of the Fisher information matrix. The update rule is modified to
\begin{equation}
    {\vec{\theta}}^{(k+1)} = {\vec{\theta}}^{(k)} - \eta S^{-1} \nabla_{\vec{\theta}} \operatorname{Cost}(\vec{\theta}),
\end{equation}
where $\eta$ is the learning rate, $S$ is the metric tensor, and $\nabla_{\vec{\theta}} \operatorname{Cost}(\vec{\theta})$ is the gradient of the energy functional with respect to the parameters $\vec{\theta}$. 

The stochastic gradient \(\nabla_{\vec{\theta}} \operatorname{Cost}(\vec{\theta})\) is estimated over the probability distribution defined by the entries of the vectorized density matrix \(P_{\vec{\theta}}(\valpha, \vbeta) \propto |\rho_{\vec{\theta}}(\valpha, \vbeta)|^2\) as in Sec.\ref{SM:expectation}. The gradient of the cost function with respect to the complex conjugate of the \(i\)th parameter can be expressed as
\begin{equation}
    \frac{\partial}{\partial \theta_i^*} \operatorname{Cost}(\vec{\theta})
    = \langle \tilde{\mathcal{L}}_i \nabla_i^* \tilde{\mathcal{L}}_i \rangle - \langle O_i^*\,\tilde{\mathcal{L}}^2 \rangle,
\end{equation}
where the local estimator is defined by
\begin{equation}
   \tilde{\mathcal{L}}(\valpha, \vbeta) = \frac{\sum_{\valpha',\vbeta'} \hat{\mathcal{L}}(\valpha,\vbeta; \valpha', \vbeta')\,\rho_{\vec{\theta}}(\valpha',\vbeta')}{\rho_{\vec{\theta}}(\valpha, \vbeta)}.
\end{equation}

\subsubsection{Optimization Workflow}

The full optimization process is summarized as follows:

\begin{enumerate}
    \item \textbf{Initialize} the network parameters \(\vec{\theta}\) randomly.
    \item \textbf{Sample} a batch of configurations from the current transformer density operator ansatz using a Markov chain Monte Carlo sampler introduced in Sec.~\ref{SM:sample}.
    \item \textbf{Compute} the loss (cost) functional \(\operatorname{Cost}(\vec{\theta})\) and its gradient.
    \item \textbf{Compute} the metric tensor \(S\) for stochastic reconfiguration introduced in Sec.~\ref{SM:metric}.
    \item \textbf{Regularize} the metric tensor with a small constant \(\lambda\), i.e., \(S' = S + \lambda \, \identity\), to ensure numerical stability as introduced in Sec.~\ref{SM:SR}.
    \item \textbf{Solve} for the preconditioned gradient update using \(S'^{-1} \nabla_{\vec{\theta}} \operatorname{Cost}(\vec{\theta})\).
    \item \textbf{Update} the parameters \(\vec{\theta}\) using the rule
    \[
    {\vec{\theta}}^{(k+1)} = {\vec{\theta}}^{(k)} - \eta S'^{-1} \nabla_{\vec{\theta}} \operatorname{Cost}(\vec{\theta}).
    \]
    \item \textbf{Repeat} the above steps until convergence.
\end{enumerate}

This comprehensive optimization strategy—integrating first-order gradient updates, second-order corrections via Stochastic Reconfiguration, and the conceptual framework of NetKet’s steady-state variational driver—ensures that our transformer density operator ansatz accurately converges to the steady state of open quantum systems.

\subsubsection{Learning Rate Scheduling}
\label{SM:sche}
To improve training stability and convergence, we employ a learning rate schedule that combines an initial warm-up step with a cosine decay strategy. Specifically, the learning rate is defined as:
\begin{equation}
    \mathrm{lr}(i_{\mathrm{step}}) =
    \begin{cases}
        \eta_0, & i_{\mathrm{step}} < i_{\mathrm{switch}} \\
        \eta_0 \cdot \frac{1 + \cos(\pi (i_{\mathrm{step}} - i_{\mathrm{switch}}) / i_{\mathrm{decay}})}{2} + \alpha \eta_0, & i_{\mathrm{step}} \geq i_{\mathrm{switch}}
    \end{cases}
\end{equation}
where \(\eta_0\) is the initial learning rate, \(i_{\mathrm{step}}\) is the current training step, \(i_{\mathrm{switch}}\) denotes the step at which the decay begins, \(i_{\mathrm{decay}}\) is the decay period, and \(\alpha\) is a scaling factor for the minimum learning rate. In our implementation, we set:
\begin{align*}
    \eta_0 &= 0.0061, \quad i_{\mathrm{switch}} = 30000, \quad i_{\mathrm{decay}} = 40000, \quad \alpha = 0.001.
\end{align*}
This schedule ensures a stable learning rate during the initial phase, facilitating rapid exploration of the parameter space, followed by a smooth decay to refine the variational ansatz.

Additionally, we apply a similar schedule to the stochastic reconfiguration preconditioner, adjusting the diagonal shift dynamically to improve numerical stability and convergence:
\begin{widetext}
\begin{equation}
    \lambda_{\mathrm{SR}}(i_{\mathrm{step}}) =
    \begin{cases}
        \lambda_0, & i_{\mathrm{step}} < i_{\mathrm{switch,SR}} \\
        \lambda_0 \cdot \frac{1 + \cos\left[\pi \left(i_{\mathrm{step}} - i_{\mathrm{switch,SR}}\right) / i_{\mathrm{decay,SR}}\right]}{2} + \alpha_{\mathrm{SR}} \lambda_0, & i_{\mathrm{step}} \geq i_{\mathrm{switch,SR}}
    \end{cases}
\end{equation}
\end{widetext}
where \(\lambda_0 = 0.004\), \(i_{\mathrm{switch,SR}} = 30000\), \(i_{\mathrm{decay,SR}} = 40000\), and \(\alpha_{\mathrm{SR}} = 0.01\). This approach dynamically adjusts the regularization strength of the stochastic reconfiguration method, ensuring robustness while maintaining efficiency in parameter updates.

By incorporating these schedules, we balance initial exploration with controlled optimization, leading to improved stability and convergence of the transformer density operator ansatz.

\subsection{Benchmark Methods for Steady-State Computation}
\label{SM:Ben}

To validate our variational approach, we compare the steady-state observables computed with our transformer-based ansatz against two benchmark methods implemented by NetKet. Both methods aim to solve for the steady state of an open quantum system, which satisfies
\begin{equation}
    \hat{\mathcal{L}} |\rho\rrangle = 0,
\end{equation}
where \(\hat{\mathcal{L}}\) is the Lindblad superoperator.

\subsubsection{Exact Diagonalization}
For small system sizes of less than 7 spins, one can fully diagonalize the Lindblad superoperator. In the realization, the following operator is constructed 
\begin{equation}
    \mathfrak{L} = \hat{\mathcal{L}}^{\dagger} \hat{\mathcal{L}}.
\end{equation}
Exact diagonalization proceeds by solving the eigenvalue problem
\begin{equation}
    \mathfrak{L} |\rho\rrangle = \lambda |\rho\rrangle.
\end{equation}
The steady state is identified as the eigenvector corresponding to the zero eigenvalue (\(\lambda=0\)):
\begin{equation}
    \mathfrak{L} |\rho_{\mathrm{ss}}\rrangle = 0.
\end{equation}

\subsubsection{Iterative Biconjugate Gradient Stabilized Method}

For larger system sizes, the Hilbert space grows exponentially, making full diagonalization computationally infeasible. To efficiently obtain the steady state in such cases, we employ the iterative Biconjugate Gradient Stabilized (BiCGStab) method. As before, we seek to solve  
\begin{equation}
    \hat{\mathcal{L}}^{\dagger}\hat{\mathcal{L}} \,|\rho\rrangle = \mathfrak{L} \,|\rho\rrangle = 0 ,
\end{equation}  
where \(\mathfrak{L}\) is a positive semi-definite Hermitian operator. The BiCGStab algorithm allows us to iteratively converge to the steady-state solution without requiring explicit matrix inversion or full diagonalization, making it particularly suitable for large-scale dissipative quantum systems. \rev{For instance, computing the steady state of a 13-site system using this iterative method requires substantial computational resources and time, typically taking several days even with optimized implementations, highlighting the computational advantages of our transformer-based approach}.

\vspace{1em}
\noindent
\textbf{Residual and Krylov Subspace.}
For a given approximate solution \(|\rho^{(k)}\rrangle\) at iteration \(k\), the \emph{residual} is defined as
\begin{equation}
    |r^{(k)}\rangle = \mathfrak{L} \, |\rho^{(k)}\rrangle.
\end{equation}
The BiCGStab algorithm constructs approximate solutions by searching within the so-called \emph{Krylov subspace} generated by repeatedly applying \(\mathfrak{L}\) to the initial residual. Concretely, starting from the initial residual \(\,|r^{(0)}\rangle\), the Krylov subspace of dimension \(m\) is given by
\begin{equation}
    \mathcal{K}_m\bigl(\mathfrak{L}, |r^{(0)}\rangle\bigr) 
    = \mathrm{span}\Bigl\{ \,|r^{(0)}\rangle,\, \mathfrak{L}|r^{(0)}\rangle,\, \mathfrak{L}^2|r^{(0)}\rangle,\dots,\mathfrak{L}^{m-1}|r^{(0)}\rangle \Bigr\}.
\end{equation}
At each iteration, BiCGStab refines \(|\rho^{(k)}\rrangle\) within this subspace to reduce the norm of the residual \(\|\,|r^{(k)}\rangle\|\).

\vspace{1em}
\noindent
\textbf{Algorithmic Steps.}
\begin{enumerate}
    \item \textbf{Initialization:} 
    Choose an initial guess \(|\rho^{(0)}\rrangle\). Compute the initial residual \(|r^{(0)}\rangle = \mathfrak{L}|\rho^{(0)}\rrangle\). Often, one sets \(|\rho^{(0)}\rangle\) to a random vector or a simple ansatz.
    \item \textbf{Iteration:}
    At iteration \(k\), BiCGStab updates \(|\rho^{(k)}\rrangle\) by forming a new approximation \(|\rho^{(k+1)}\rrangle\) that ideally satisfies a smaller residual within a Krylov subspace:
    \begin{equation}
      |r^{(k+1)}\rangle = \mathfrak{L} \, |\rho^{(k+1)}\rrangle.
    \end{equation}
    The method employs additional auxiliary vectors (e.g., search directions and a “shadow” residual) to stabilize convergence and avoid breakdowns inherent in BiCGStab.
    \item \textbf{Convergence:} 
    Once the norm of the residual \(\bigl\|\,|r^{(k)}\rangle\bigr\|\) is smaller than a prescribed tolerance (e.g., \(\epsilon = 10^{-7}\)), the current approximation \(|\rho^{(k)}\rrangle\) is taken as the steady state:
    \[
       |\rho_{\mathrm{ss}}\rrangle \equiv |\rho^{(k)}\rrangle.
    \]
\end{enumerate}

By constructing and updating these Krylov subspace approximations, BiCGStab efficiently converges to the zero-eigenvalue solution of \(\mathfrak{L} = \hat{\mathcal{L}}^{\dagger}\hat{\mathcal{L}}\), even in high-dimensional spaces.

Once the steady state \(|\rho_{\mathrm{ss}}\rrangle\) is obtained, the expectation value of an observable \(\hat{O}\) is computed via
\begin{equation}
    \langle \hat{O} \rangle = \frac{\Tr(\hat{O}\hat{\rho}_{\mathrm{ss}})}{\Tr(\hat{\rho}_{\mathrm{ss}})},
\end{equation}
or, equivalently in the vectorized notation,
\begin{equation}
    \langle \hat{O} \rangle = \frac{\llangle \hat{O} | \hat{\rho}_{\mathrm{ss}} \rrangle}{\llangle \identity | \hat{\rho}_{\mathrm{ss}} \rrangle}.
\end{equation}

This approach enables us to compute observables without explicitly constructing the full Hilbert space, making it well-suited for large systems as a baseline.

\subsection{Evaluation of the Optimized Ansatz}
\label{SM:eval}
After training our transformer density operator ansatz, we obtain an optimized parameter set \(\vec{\theta}\) that approximates the steady-state density operator. To assess the accuracy of our transformer-based density operator ansatz, we compute the expectation values of local observables via Monte Carlo sampling, as discussed in Sec.~\ref{SM:sample}.

For example, consider the spatial average of the Pauli operator \(\sigma^z\),
\begin{equation}
    \langle \sigma^z \rangle = \frac{1}{N} \sum_{i=1}^N \langle \sigma^z_i \rangle.
\end{equation}
There are two ways to obtain $\langle \sigma^z_i \rangle$: variational approach and exact approach.

\rev{
\begin{figure*}[th]
\centering
\includegraphics[width=1\textwidth]{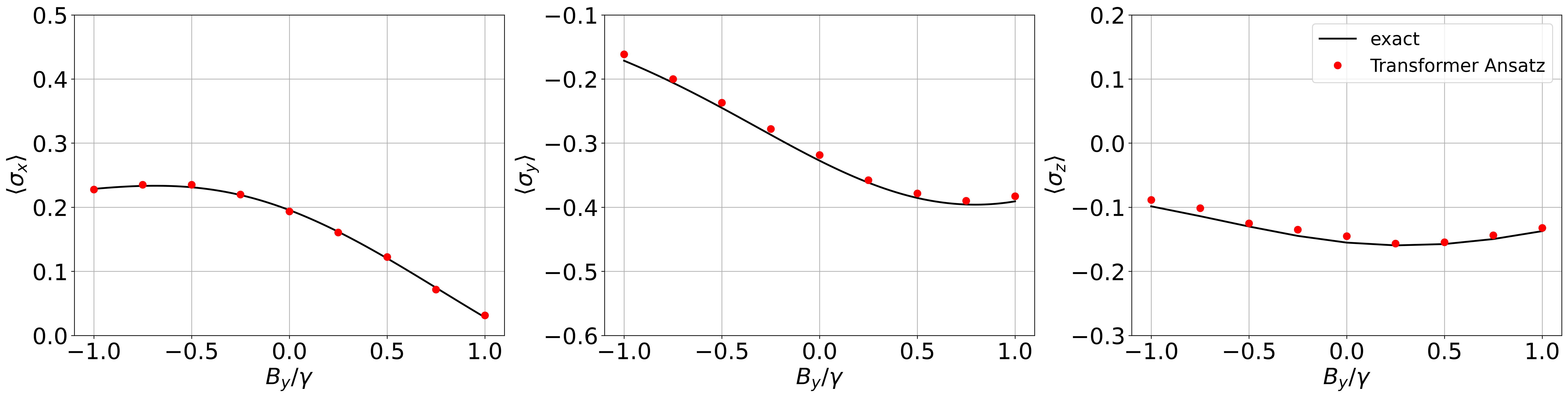}
\caption{ \rev{Expectation values of the averaged magnetization components $\langle \sigma_x \rangle$, $\langle \sigma_y \rangle$, and $\langle \sigma_z \rangle$ as functions of the transverse magnetic field $B_y / \gamma$ for a dissipative Heisenberg spin chain with $N = 10$ sites. The system parameters are $J_x / \gamma = 1.4$, $J_y / \gamma = 2.0$, $J_z / \gamma = 1.0$, $B_x / \gamma = -1.0$, and $B_z / \gamma = 0.1$, with periodic boundary conditions and uniform dissipation rate $\gamma$. The exact expectation values (black lines) were obtained using the iterative BiCGStab solver via NetKet, while the red dots represent results from our transformer density operator ansatz.}
}
\label{fig:Fig_heisen_10}
\end{figure*}
}

In the variational approach, the expectation value of an arbitrary operator \(\hat{O}\) is computed as
\begin{equation}
    \langle \hat{O} \rangle = \frac{\Tr\bigl(\hat{\rho}_{\vec{\theta}} \hat{O}\bigr)}{\Tr\bigl(\hat{\rho}_{\vec{\theta}}\bigr)}.
\end{equation}
This can be recast as a classical expectation value over a probability distribution defined on the diagonal elements of \(\rho_{\vec{\theta}}\):
\begin{equation}
    \langle \hat{O} \rangle = \sum_{\valpha} P_{\vec{\theta}}(\valpha)\, \tilde{O}_{\rho_{\vec{\theta}}}(\valpha),
\end{equation}
where the probability distribution over diagonal elements of the density matrix is
\begin{equation}
\label{eq:prob}
    P_{\vec{\theta}}(\valpha) = \frac{\rho_{\vec{\theta}}(\valpha, \valpha)}{\Tr\bigl(\hat{\rho}_{\vec{\theta}}\bigr)},
\end{equation}
and the local estimator \(\tilde{O}_{\rho_{\vec{\theta}}}(\valpha)\) is defined by
\begin{equation}
    \tilde{O}_{\rho_{\vec{\theta}}}(\valpha) = \sum_{\vbeta} \frac{\rho_{\vec{\theta}}(\valpha,\vbeta)}{\rho_{\vec{\theta}}(\valpha, \valpha)} \langle \vbeta | \hat{O} | \valpha \rangle.
\end{equation}
For \(\hat{O} = \sigma^z_i\) this becomes
\begin{equation}
    \tilde{\sigma}^z_{\rho_{\vec{\theta}}}(\valpha) = \sum_{\vbeta} \frac{\rho_{\vec{\theta}}(\valpha,\vbeta)}{\rho_{\vec{\theta}}(\valpha,\valpha)} \langle \vbeta | \sigma^z_i | \valpha \rangle,
\end{equation}
so that
\begin{equation}
    \langle \sigma^z_i \rangle = \sum_{\alpha} P_{\vec{\theta}}(\valpha)\, \tilde{\sigma}^z_{\rho_{\vec{\theta}}}(\valpha).
\end{equation}

There is another way to calculate the expected value
\begin{equation}
    \langle \sigma^z_i \rangle = \Tr\bigl(\sigma^z_i \hat{\rho}\bigr) = \llangle \sigma^z_i | \rho \rrangle,
\end{equation}
with
\begin{equation}
    \llangle \sigma^z_i | \rho \rrangle = \sum_{\valpha,\vbeta} \rho_{\valpha,\vbeta}\, \langle \sigma^z_i | \valpha,\vbeta \rangle,
\end{equation}
and the identification
\begin{equation}
    \langle \sigma^z_i | \valpha,\vbeta \rangle \equiv \langle \vbeta | \sigma^z_i | \valpha \rangle.
\end{equation}
If one could sum over the entire Hilbert space, this method would yield exact expectation values. However, for systems with a large number of particles, the Hilbert space grows exponentially, making such a full summation computationally infeasible. Therefore, for large systems, we rely on Monte Carlo sampling methods, which offer an efficient and approximate approach to evaluating observables without the need to compute the full density matrix.

In summary, after optimizing the ansatz, we compute observables (e.g., \(\langle \sigma^x \rangle\), \(\langle \sigma^y \rangle\), and \(\langle \sigma^z \rangle\)) by sampling from \(P_{\vec{\theta}}(\valpha)\) in Eq.~\eqref{eq:prob} and evaluating the corresponding local estimators as described above. We compare our results with baselines calculated using NetKet (see Sec.~\ref{SM:Ben}). For small system sizes (\(N < 7\)), we employ exact diagonalization of the full Lindblad superoperator. For larger system sizes, we use the iterative BiCGStab method, which directly solves the steady-state equation \(\hat{\mathcal{L}}{\rho} = 0\). The excellent agreement between these measurements under our optimized ansatz and benchmark solutions (obtained via exact diagonalization for small systems or iterative solvers for larger systems) confirms that our transformer-based ansatz accurately captures the steady-state properties of open quantum systems.

\rev{
\section{More numerical results}
\label{sup:more}

\rev{
\begin{figure*}[th]
\centering
\includegraphics[width=1\textwidth]{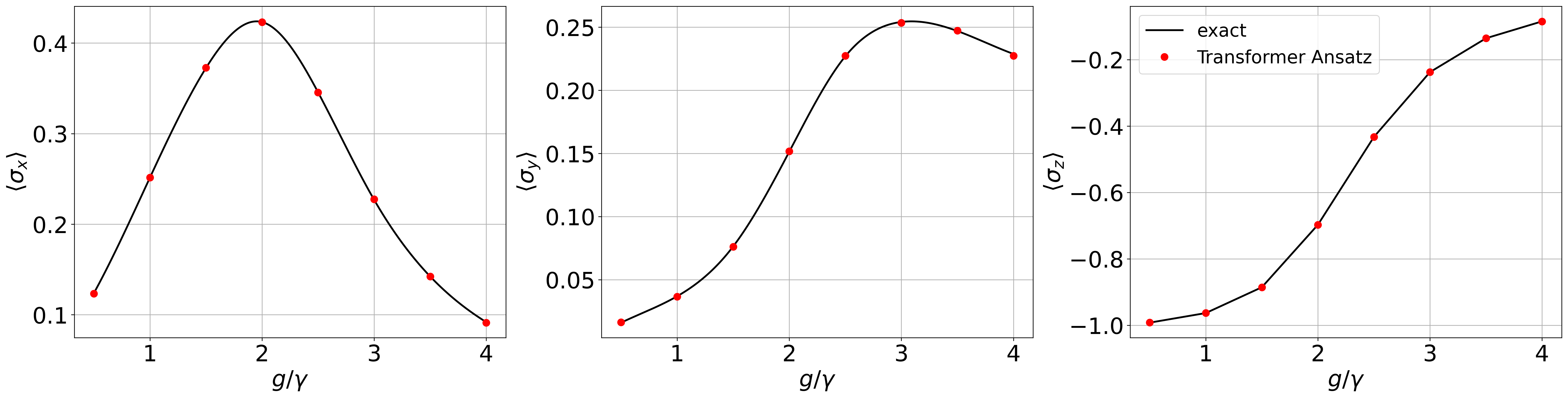}
\caption{ \rev{Expectation values of the averaged magnetization components $\langle \sigma_x \rangle$, $\langle \sigma_y \rangle$, and $\langle \sigma_z \rangle$ as functions of the normalized transverse field strength $g/\gamma$ for a 2D dissipative Ising model on a $2 \times 2$ square lattice with periodic boundary conditions. The system parameters are $V = 2\gamma$ with uniform dissipation rate $\gamma$ across all 4 sites. Our transformer density operator ansatz (red dots) accurately captures the system's behavior obtained by NetKet.}
}
\label{fig:2d22_ising}
\end{figure*}
}

\emph{The 1D Heisenberg Model with larger number of spins.} ---
To further demonstrate the scalability and robustness of our transformer density operator ansatz, we present additional results for a dissipative Heisenberg spin chain with $N=10$ sites. The system is governed by the same Hamiltonian as described in the main text:
\begin{equation}
H = \sum_{i=1}^N \sum_{k=x,y,z} \left( J_k \sigma_i^k \sigma_{i+1}^k + B_k \sigma_i^k \right),
\end{equation}
with periodic boundary conditions and uniform dissipation rate $\gamma$ across all sites. The jump operators are chosen as $L_j = \sigma_j^-$ for local spin lowering at each site.

For this $N=10$ system, we use the same parameters as in the main text: interaction strengths $J_x / \gamma = 1.4$, $J_y / \gamma = 2.0$, and $J_z / \gamma = 1.0$, with magnetic field components $B_x / \gamma = -1.0$ and $B_z / \gamma = 0.1$. We vary the transverse magnetic field $B_y / \gamma$ to explore different regimes of the system's behavior. The exact benchmark results are obtained using the iterative BiCGStab solver implemented in NetKet, which provides high-accuracy solutions for systems of this size.

As shown in Fig.~\ref{fig:Fig_heisen_10}, our transformer density operator ansatz maintains good accuracy across the entire parameter range for the $N=10$ system. This result confirms the scalability and reliability of our approach across different system sizes within the one-dimensional Heisenberg model family.

\vspace{1em}
\emph{Two-dimensional Ising Model on a $2\times 2$ Lattice.} ---
To provide additional validation of our approach for 2D systems and demonstrate consistency across different lattice sizes, here we present results for a smaller $2 \times 2$ square lattice dissipative Ising model.

The Hamiltonian for this system follows the same form as described in the main text:
\begin{equation}
H = \frac{V}{4} \sum_{\langle i j \rangle} \sigma^z_i \sigma^z_j + \frac{g}{2} \sum_{i=1}^{4} \sigma^x_i,
\end{equation}
where the sum over nearest neighbors $\langle i j \rangle$ now includes the 4 bonds in the $2 \times 2$ lattice with periodic boundary conditions. The dissipative dynamics are governed by local jump operators $L_i = \sigma^-_i$ with uniform dissipation rate $\gamma$ across all sites, and we set the interaction strength to $V = 2\gamma$.

The results presented in Fig.~\ref{fig:2d22_ising} show the expectation values of the averaged magnetization components as functions of the normalized transverse field strength $g/\gamma$. Our transformer density operator ansatz (red dots) shows excellent agreement with the exact results (black lines) across the entire parameter range.

}

\bibliographystyle{apsrev4-1-title}
\bibliography{mybib}

\end{document}